\documentclass[prd,showpacs,superscriptaddress,notitlepage,longbibliography]{revtex4-1}
\usepackage[usenames,dvipsnames]{color}
\usepackage{amsmath}
\usepackage{amssymb}
\usepackage{latexsym}
\usepackage{enumerate}
\usepackage{bbm}
\usepackage{amsthm}
\usepackage[colorlinks]{hyperref}
\usepackage{graphicx}
\usepackage{dcolumn}
\usepackage{bm}
\hypersetup{citecolor=red, linkcolor=blue}
\usepackage{subfigure}
\usepackage{footnote}
\usepackage{ulem}
\usepackage{mathrsfs}
\DeclareSymbolFontAlphabet{\mathrsfs}{rsfs}
\newcommand{\scri}{\mathrsfs{I}}
\newcommand{\const}{\mathrm{const}}
\renewcommand{\emph}[1]{\textit{#1}}
\newtheorem*{conjecture}{Conjecture}
\begin{document}

\title{Numerical construction of initial data for Einstein's equations\\
  with static extension to space-like infinity}

\author{Georgios Doulis}
\affiliation{Max Planck Institute for Gravitational Physics 
  (Albert Einstein Institute), Am M\"uhlenberg 1, 14476 Potsdam, Germany}

\author{Oliver Rinne}
\affiliation{Max Planck Institute for Gravitational Physics 
  (Albert Einstein Institute), Am M\"uhlenberg 1, 14476 Potsdam, Germany}
\affiliation{Department of Mathematics and Computer Science,
  Freie Universit\"at Berlin, Arnimallee 2-6, 14195 Berlin, Germany}

\date{\today} 

\vspace{1cm}

\begin{abstract}
  We describe a numerical method to construct Cauchy data extending to
  space-like infinity based on Corvino's (2000) gluing method.
  Adopting the setting of Giulini and Holzegel (2005), we restrict ourselves 
  here to vacuum axisymmetric spacetimes and glue a Schwarzschildean end to 
  Brill-Lindquist data describing two non-rotating black holes.
  Our numerical implementation is based on pseudo-spectral methods, and we carry
  out extensive convergence tests to check the validity of our numerical results.
  We also investigate the dependence of the total ADM mass on the details of 
  the gluing construction. 
\end{abstract}

\maketitle

\numberwithin{equation}{section}


\section{Introduction}
\label{sec:intro}

Many situations of astrophysical interest can be described to good approximation
as isolated systems: an asymptotically flat spacetime containing a compact
self-gravitating source such as a collapsing star, a black hole binary, etc.
A fundamental problem in the numerical solution of the Einstein equations for
such systems is the treatment of the far field.
Access to the asymptotic region known as conformal 
infinity \cite{FrauendienerLRR} is important for several reasons.
Firstly, gravitational radiation is only defined in an unambiguous way at future
null infinity.
Including this region in the computational domain enables extraction of the
gravitational radiation emitted by the source in a straightforward way.
This is important for the modelling of astrophysical sources of gravitational
radiation.
Secondly, many open problems in mathematical relativity such as black hole
stability and cosmic censorship are statements about the global structure of
spacetime.
If numerical studies are to shed light on these questions then access to 
conformal infinity is indispensable.

The standard approach to numerical relativity is based on the Cauchy 
formulation of Einstein's equations.
The $t=\const$ slices are truncated at a finite distance from the source, 
where boundary conditions are imposed.
These must ensure that the resulting initial-boundary value problem is
well posed, they must be compatible with the constraints that hold on the
individual $t=\const$ slices, and ideally they should be absorbing, 
i.e.~the artificial boundary should be transparent to gravitational radiation.
Despite much progress in this direction (see \cite{SarbachLRR} for a 
review article), this approach is necessarily limited because exact 
absorbing boundary conditions cannot be defined at a finite distance in the
full nonlinear theory of general relativity so that linearisation about
a given background spacetime is typically assumed.
And imperfect boundary conditions can easily destroy relevant features of the
solutions such as late-time power-law tails caused by the backscattering of 
gravitational radiation.

\begin{figure}[htb]
 \centering
 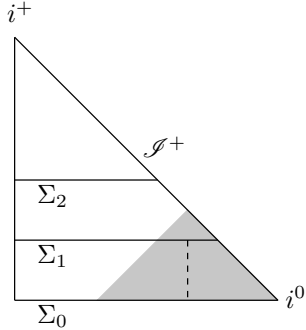
 \caption{
   Penrose diagram illustrating the use of our initial data for the evolution problem. 
   The data on the Cauchy surface $\Sigma_0$ are glued to a Schwarzschild end
   (shaded region) and evolved to the surface $\Sigma_1$ using Cauchy evolution
   with a finite boundary (dashed line) in the Schwarzschild region.
   Given the known Schwarzschild solution, the data on $\Sigma_1$ can be
   completed to a hyperboloidal surface reaching $\scri^+$.
   This then serves as initial data for a hyperboloidal evolution code. 
 }
 \label{fig:evolution}
\end{figure}

An alternative to evolution on truncated Cauchy slices is evolution on 
hyperboloidal slices extending to future null infinity $\scri^+$.
(Examples of hyperboloidal slices are the slices $\Sigma_1$ and $\Sigma_2$
in Fig.~\ref{fig:evolution}.)
In this approach a conformal transformation is applied to the spacetime metric,
combined with a compactifying coordinate transformation that maps infinity
to a finite location.
The conformal boundary of the slices becomes a pure outflow boundary so that
no boundary conditions are required there.
Hyperboloidal evolution was first advocated in general relativity by Friedrich
in the context of his regular conformal field equations \cite{Friedrich1983a},
a symmetric hyperbolic formulation of the (suitably augmented) Einstein
equations that is completely regular up to the conformal boundary.
For reviews of the theoretical development as well as numerical implementations 
based on this system, see e.g.~\cite{FrauendienerLRR,Husa2002,Husa2003}.
An alternative method is based on a straightforward ADM \cite{Arnowitt1962}
split of the conformally transformed Einstein equations on hyperboloidal
surfaces of constant mean curvature \cite{Moncrief2009}.
The resulting equations are formally singular at $\scri^+$ but can nevertheless
be evaluated there in terms of regular conformal data.
Based on this system, stable numerical evolutions of a gravitationally 
perturbed Schwarzschild black hole in axisymmetry were
achieved \cite{Rinne2010}; later matter fields were also 
included \cite{Rinne2013,Rinne2014b}.
Further proposals for hyperboloidal evolution systems that, as far as we know, 
have not been implemented numerically yet can be found 
in \cite{Zenginoglu2008,Bardeen2011}.

The hyperboloidal surfaces are only partial (in our case, future) Cauchy surfaces. 
The problem remains how to evolve entire spacetimes from Cauchy data extending 
to space-like infinity. The main difficulty here is that part of the Cauchy data---namely 
some of the components of the Weyl tensor---are singular at space-like infinity 
if the ADM mass is not zero \cite{Friedrich1988}. In \cite{Friedrich1998} Friedrich 
proposed a way to render these Cauchy data regular while guaranteeing the regularity 
of the conformal field equations at space-like infinity. The basic ingredient of 
this approach is the blowing up of space-like infinity $i^0$ to a \textit{cylinder} 
$I = [-1, 1] \times \mathbb{S}^2$ that serves as a link of finite length (along 
the time direction) between past $\scri^-$ and future $\scri^+$ null infinity. 
The 2-spheres $I^\pm = I \cap \scri^\pm$ where the cylinder meets future and past 
null infinity are called \textit{critical sets}. The equations that propagate the 
data from $\scri^-$ to $\scri^+$ along the cylinder acquire an extremely simple 
form in Friedrich's representation that makes them ideal for numerical implementation, 
see \cite{Beyer2012,Doulis2013,Beyer2014a,Beyer2014b,Frauendiener2014} for some 
recent numerical work. On the cylinder all the spatial derivatives drop out. Therefore, 
the cylinder is a \textit{total characteristic} of the system and hence no boundary 
conditions are required there. However, this intrinsic system of propagation equations 
degenerates at the critical sets $I^\pm$ and develops logarithmic singularities 
there that are expected to travel along null infinity and spoil its smoothness. 
In Friedrich's approach this generic singular behaviour is successfully reproduced. 
Its appearance has been made explicit and related to the structure of the initial 
data. In other words, there is a possibility that by choosing appropriately the 
initial data the occurrence of non-smooth features in the solutions at null infinity 
can be avoided. A possible solution proposed already in \cite{Friedrich1998} is 
to prescribe initial data that respect a set of regularity conditions involving 
the Cotton tensor. However it turned out \cite{Kroon2004} that these conditions 
are not sufficient to prevent the occurrence of the logarithmic singularities in 
higher order expansions of the solutions of the intrinsic system of propagation 
equations. In \cite{Kroon2004} Valiente Kroon proposed a new regularity condition 
in the form of the following conjecture:
\begin{conjecture}
 If an initial data set which is time symmetric and conformally flat in a neighbourhood
 of infinity yields a development with a smooth null infinity, then the initial 
 data is in fact Schwarzschildean in that neighbourhood.
\end{conjecture}
Recently, the results in \cite{Kroon2010, Kroon2012} have pointed in favour of 
the conjecture, but there is still work to be done in order to fully prove it. 
What has been shown is that the solution is smooth at the critical sets if and 
only if the initial data is exactly Schwarzschildean in a neighbourhood of infinity. 
It remains to be proved that the development of the solution along null 
infinity is smooth if and only if it is smooth at the critical sets. 
If true, the conjecture unveils the special role that static data play 
in the smooth development of Cauchy data extending to space-like infinity.

One might object that initial data that are static in a neighbourhood of
space-like infinity are overly restrictive.
However, a powerful result by Corvino \cite{Corv:2000} suggests that this is
not the case.
He showed that any given asymptotically flat and conformally flat initial data 
can be truncated and glued along an annulus to a Schwarzschild metric 
in the exterior, provided the radius of the gluing annulus is sufficiently
large and the mass of the exterior Schwarzschild metric is chosen appropriately.
There are otherwise no additional restrictions on the metric in the interior, in 
particular non-static spacetimes including gravitational radiation are allowed.
The method has been generalised to stationary rotating ends described 
by the Kerr metric, and a cosmological constant has been included 
\cite{Corvino2006,Chrusciel2008,Chrusciel2009,Cortier2013}.

Corvino's result can be used for the evolution problem as follows
(see also \cite{Chrusciel2002}). 
Since his initial data are
Schwarzschild in a neighbourhood of space-like infinity $i^0$ on the initial 
Cauchy slice $\Sigma_0$ (see Fig.~\ref{fig:evolution}),
the future development of these initial data will also be Schwarzschild
in a neighbourhood of $i^0$ (the shaded region in Fig.~\ref{fig:evolution}).
By placing an artificial timelike boundary in this region, the data on 
$\Sigma_0$ can be evolved to the future for some time using standard Cauchy 
evolution with \textit{exact} boundary conditions taken from the known
Schwarzschild solution.
From this evolution, data on a hypersurface $\Sigma_1$ are obtained, 
e.g.~a hypersurface of constant mean curvature.
Outside the artificial boundary, the solution on $\Sigma_1$ is known
analytically (Schwarzschild), so we obtain data on a complete hyperboloidal
surface.
These can then be taken as initial data for a hyperboloidal evolution code.
For the problem studied in the present paper (vacuum axisymmetric spacetimes),
the code developed in \cite{Rinne2010} can in principle be used.

The present paper deals with the first step of this proposal, namely the
construction of initial data based on Corvino's gluing method.
It should be stressed that the proof of Corvino's theorem is not explicit,
i.e.~it does not provide us with a prescription for how to actually 
construct the glued initial data.
One of the aims of this paper is to compute such data numerically,
at least in a simple setting.
We assume here that spacetime is vacuum and axisymmetric.
Corvino's method under these assumptions was first studied analytically by
Giulini and Holzegel \cite{Giu&Hol:2005}.
An important achievement of this paper was to turn Corvino's idea into 
an explicit PDE problem that can, in principle at least, be solved to obtain 
the glued data.
The 3-metric at a moment of time symmetry in a vacuum axisymmetric spacetime
can be written in the form of a Brill wave \cite{Brill:1959}.
This comprises both the Schwarzschild solution in isotropic coordinates and,
by superposition, Brill-Lindquist data \cite{Bril&Linq:1963} for an axisymmetric
configuration of two non-rotating black holes (not in equilibrium).
Giulini and Holzegel took the metric in the interior to be Brill-Lindquist
and glued it to a Schwarzschild metric in the exterior using a general
Brill wave metric on the gluing annulus.
They were mainly interested in the question whether the ADM mass (i.e., the
mass of the exterior Schwarzschild solution) can be smaller than the sum of
the two Brill-Lindquist black hole masses, as they expected that this would 
reduce the (generally unwanted) gravitational radiation introduced in the 
gluing region. They claimed that this can be done at least to first order in 
the inverse gluing radius. Using numerical methods we are able to study the 
solution also for smaller gluing radii.

This paper is organised as follows.
In Sec.~\ref{sec:gluing_construction} we describe the details of the gluing
construction and derive the equations to be solved.
A novel ingredient is an integrability condition that fixes the relation 
between the masses of the Brill-Lindquist black holes and the exterior
Schwarzschild solution (Sec.~\ref{sec:integrability}).
Sec.~\ref{sec:numer_implement} is devoted to the numerical implementation.
We describe the pseudo-spectral method we use (Sec.~\ref{sec:numer_scheme}) and
test the code with an artificial exact solution (Sec.~\ref{sec:exact_sol})
before turning to the actual gluing problem in 
Sec.~\ref{sec:numer_gluing_constr}.
Detailed convergence tests are carried out.
Finally, we investigate how the total ADM mass depends on the details of the 
gluing procedure (Sec.~\ref{sec:reduction}).
We conclude with a discussion of our results and an outlook on future work 
in Sec.~\ref{sec:discussion}.


\section{The gluing construction}
\label{sec:gluing_construction}

Following the line of thought in \cite{Giu&Hol:2005}, we set up here the mathematical framework 
on which our numerical study of the gluing construction in the subsequent section will be 
based. We will also derive an integrability condition that unveils the dependence of the 
ADM mass on the details of the gluing construction.

\subsection{Basic ingredients}
\label{sec:basics}

\begin{figure}[htb]
 \centering
 \includegraphics[height=6cm]{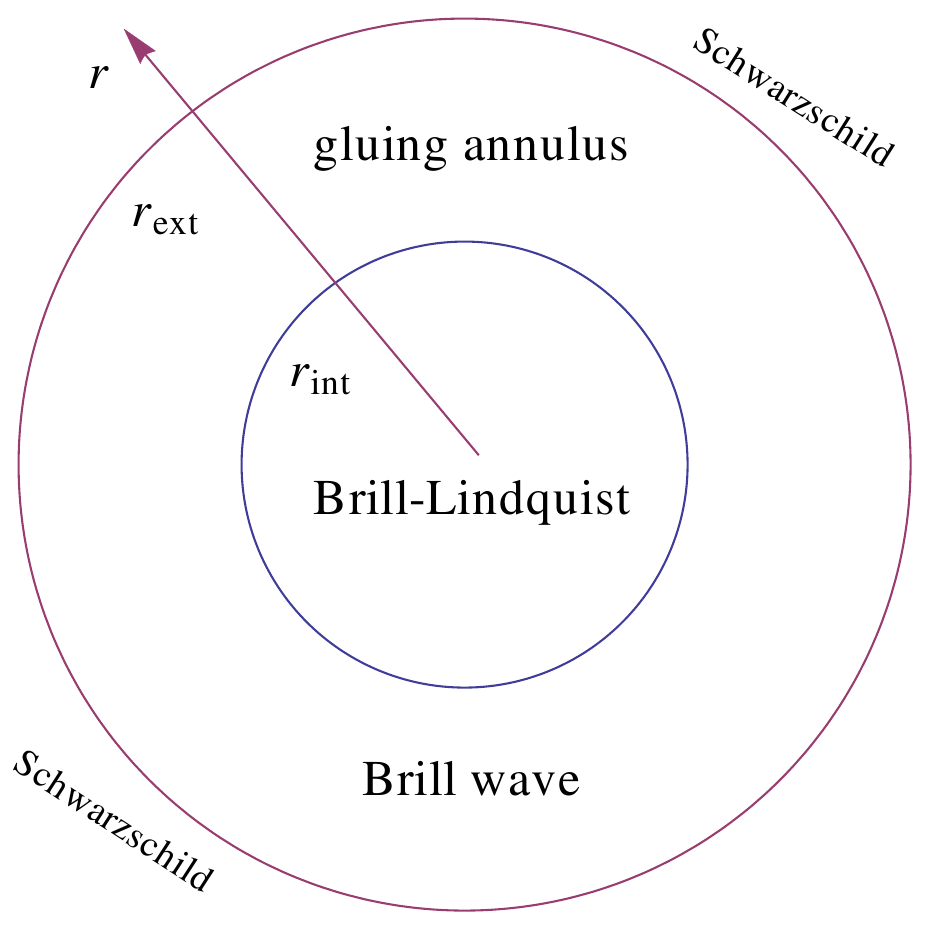}
 \caption{Graphical two-dimensional representation of the gluing construction we are going to 
 consider in the following. A Schwarzschildean end will be glued to Brill-Lindquist data 
 representing two non-rotating black holes in the interior along a transition region equipped 
 with a Brill wave metric.
 }
 \label{fig:2D_gluing}
\end{figure}

Fig.~\ref{fig:2D_gluing} encapsulates the basic features of the construction proposed in 
\cite{Giu&Hol:2005}: the interior spacetime consists of Brill-Lindquist data, the exterior 
spacetime extending to space-like infinity is Schwarzschild, and the transition between 
the two data sets takes place along a gluing annulus which is equipped with a Brill wave 
metric. The gluing annulus extends from $r_{\mathrm{int}}$ to $r_{\mathrm{ext}}$.

\begin{figure}[htb]
 \centering
 \includegraphics[height=6cm]{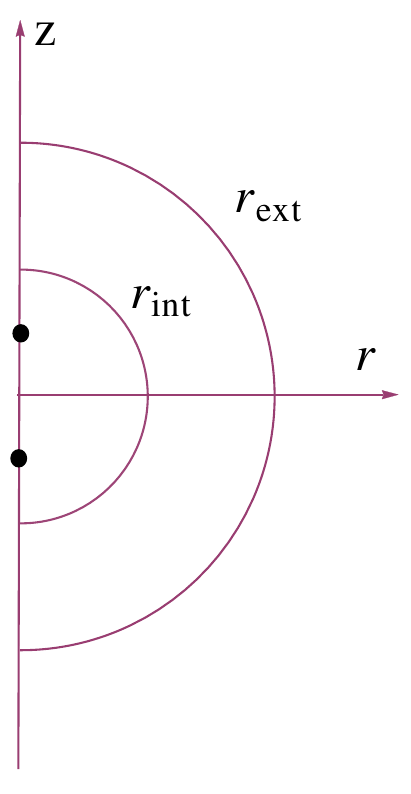}
 \caption{The Brill-Lindquist data. We consider two black holes of equal mass positioned 
 symmetrically to the origin on the z-axis. The two black holes are at a moment of time 
 symmetry, i.e. they do not carry any spin and they are momentarily at rest relative to 
 each other.}
 \label{fig:B-L_data}
\end{figure}

More specifically, in the interior ($r \leq r_{\mathrm{int}}$) we consider axisymmetric 
vacuum Brill-Lindquist data \cite{Bril&Linq:1963} describing two black holes at a moment 
of time symmetry, 
\begin{equation}
 \label{B-L_metric1}
 g_{\textrm{\tiny B-L}} = \left(1 + \frac{m_1}{2|\vec{r} - \vec{c}_1|} + \frac{m_2}{2|\vec{r} - \vec{c}_2|} \right)^4 \delta,
\end{equation}
where $\delta = dr^2 + r^2 (d\theta^2 + \sin^2\theta\, d\phi^2)$ denotes the three-dimensional 
Euclidean line element in spherical polar coordinates, and $m_k$ and $\vec c_k$, with $k = 1,2$, 
are the bare masses and coordinate centres of the two black holes, respectively. In order 
to simplify our formulation, we will assume in the following that the two black holes are 
of equal mass, i.e. $m_1 = m_2 = m$, and that they lie symmetrically to the origin on the 
z-axis, i.e. $ \vec{c}_1 = - \vec{c}_2 =  \vec{c} = (0, 0, \frac{d}{2})$, see Fig.~\ref{fig:B-L_data}. 
With these choices the line element \eqref{B-L_metric1} reduces to
\begin{equation}
 \label{B-L_metric}
 g_{\textrm{\tiny B-L}} = \left(1 + \frac{m}{2|\vec{r} - \vec{c}|} + \frac{m}{2|\vec{r} + \vec{c}|} \right)^4 \delta.
\end{equation}
Notice that the above line element is written in conformally flat form, a feature that will 
play a key role in the subsequent development of the gluing construction. It can be readily 
confirmed that the ADM mass of the Brill-Lindquist data \eqref{B-L_metric} is equal to $2m$. 

In the present work we will consider only Brill-Lindquist data where the horizons of the 
two black holes do not intersect. Also all cases where a third outer horizon \cite{Bril&Linq:1963}, 
enclosing both black holes, forms---which appears when the black holes are very close to 
each other---would not be considered here. 
 As shown in \cite{Bril&Linq:1963} both the above requirements 
are satisfied when the mass-to-distance ratio satisfies the inequality $m/d \lesssim 0.64$. 
In this setting, the radius of the event horizon of each of the black holes is given 
by the formula \cite{Bril&Linq:1963}
\begin{equation*}
 r_\mathrm{hor} = \frac{m}{2 + \frac{m}{d}}.
\end{equation*}
Therefore in the following, in order to keep the gluing annulus away from any possible horizons 
of the Brill-Lindquist data, the gluing radius $r_{\mathrm{int}}$ will be chosen in such 
a way that the inequality $r_{\mathrm{int}} > d/2 + r_\mathrm{hor}$ is always satisfied.

We intend to glue a Schwarzschildean end to the Brill-Lindquist data \eqref{B-L_metric} 
residing in the interior of our construction. Thus, in the exterior ($r \geq r_{\mathrm{ext}}$) 
of the gluing annulus we consider the usual spherically symmetric Schwarzschild data, 
which when expressed in isotropic coordinates can be written in the following conformally 
flat form,
\begin{equation}
 \label{Schw_metric}
 g_{\textrm{\tiny Schw}} = \left(1 + \frac{M}{2|\vec{r}|} \right)^4 \delta.
\end{equation}
By construction, the mass $M$ is identical with the ADM mass of the entire glued initial data.

The above two data sets \eqref{B-L_metric} and \eqref{Schw_metric} will be glued together 
using a Brill wave. This choice follows naturally from the axisymmetric 
nature of the Brill-Lindquist data considered in the interior of the construction. Brill 
waves \cite{Brill:1959} are the most general axisymmetric vacuum spacetimes with
hypersurface-orthogonal Killing vector.
In spherical coordinates, the spatial metric at a moment of time symmetry is 
given by the Weyl-type line element 
\begin{equation}
 \label{Brill_metric}
 g_{\textrm{\tiny Brill}} =  \psi^4 \left(e^{2\,q} (dr^2 + r^2 d\theta^2) + r^2 \sin^2\theta\, d\phi^2\right).
\end{equation}
The function $q(r, \theta)$ will be the unknown of our construction. 
It must satisfy the boundary conditions
\begin{equation}
 \label{bound_cond_theta}
  \begin{aligned}
   q = 0 \qquad \mathrm{for} \,\,\, \theta = 0,\, \pi, \\
   \frac{\partial q}{\partial \theta} = 0 \qquad \mathrm{for} \,\,\, \theta = 0,\, \pi.
  \end{aligned}
\end{equation}
The latter condition follows from the fact that $q$ is an even function of $\theta$. To 
justify the former, one has first to write the metric \eqref{Brill_metric} in Cartesian 
coordinates and inspect the behaviour of its metric coefficients on the z-axis; then the 
vanishing of $q$ along the z-axis follows as a necessary regularity condition that guarantees 
the absence of any conical singularities on it \cite{Rinne&Stewart:2005}. 
The conformal factor $\psi(r,\theta)$ introduced above must be positive definite everywhere and 
must satisfy the asymptotic conditions $\displaystyle \lim_{r \to \infty} \psi = 1$ at 
space-like infinity.

In summary, we want to construct a spacetime that is Brill-Lindquist \eqref{B-L_metric} 
in the interior $r \leq r_{\mathrm{int}}$, is of general Brill wave form
\eqref{Brill_metric} on the intermediate gluing annulus $r_{\mathrm{int}} \leq r \leq r_{\mathrm{ext}}$, 
and is Schwarzschild \eqref{Schw_metric} in the exterior $r \geq r_{\mathrm{ext}}$. 
In addition, all the transitions between the different regions must be smooth.

\subsection{The recipe}
\label{sec:recipe}

The novelty of Giulini's and Holzegel's construction lies in the way they incorporated 
Corvino's original idea \cite{Corv:2000} solely into the definition of the conformal factor 
$\psi$, i.e.~the metric on the entire three-dimensional time-symmetric slice is given by
the Brill wave metric \eqref{Brill_metric} with
\begin{equation}
\label{conf_factor}
 \psi = \left(1 + \frac{m}{2|\vec{r} - \vec{c}|} + \frac{m}{2|\vec{r} + \vec{c}|} \right) \beta(r,\theta) 
 +(1 - \beta(r,\theta)) \left(1 + \frac{M}{2|\vec{r}|} \right).
\end{equation}
Here $\beta(r,\theta)$ is the so-called gluing function, which apart of being smooth has the following properties:
\begin{equation}
\label{beta_function}
 \beta(r,\theta) = 
        \left\{
         \begin{array}{l l}
         1, & \quad r \leq r_{\mathrm{int}} ,\\
         0, & \quad r \geq  r_{\mathrm{ext}} ,
         \end{array} 
        \right.
\end{equation}
and all its $r$-derivatives must vanish at $r = r_\mathrm{int}$ and $r = r_\mathrm{ext}$. 
The precise form of the gluing function that is going to be used in the present work is left for Sec.~\ref{sec:numer_scheme}.

Let us see now how the gluing construction described in Sec.~\ref{sec:basics} can be realised 
by the choice \eqref{conf_factor} of the conformal factor. Notice that the first and second 
term in \eqref{conf_factor} are of Brill-Lindquist and Schwarzschildean character, respectively. 
In the interior $r \leq r_{\mathrm{int}}$ the gluing function equals unity, $\beta = 1$, 
therefore the second term in \eqref{conf_factor} vanishes. Thus, the conformal factor $\psi$ 
consists now only of its Brill-Lindquist part; inserting it into the 
Brill wave metric \eqref{Brill_metric} and enforcing $q$ to vanish in the interior region, 
the Brill wave coincides exactly with the Brill-Lindquist data 
\eqref{B-L_metric}. In a similar manner in the exterior $r \geq r_{\mathrm{ext}}$ only the 
Schwarzschildean part of $\psi$ survives, as $\beta = 0$ there. Again inserting the resulting 
conformal factor in \eqref{Brill_metric} and setting $q = 0$ also in the exterior 
region, the Brill wave \eqref{Brill_metric} coincides with 
the Schwarzschildean data \eqref{Schw_metric}. In the intermediate region $r_{\mathrm{int}} \leq r \leq r_{\mathrm{ext}}$ 
the conformal factor $\psi$, and consequently the function $q$ in \eqref{Brill_metric}, 
have a more complicated form. 

\begin{figure}[htb]
 \centering
 \includegraphics[height=7cm]{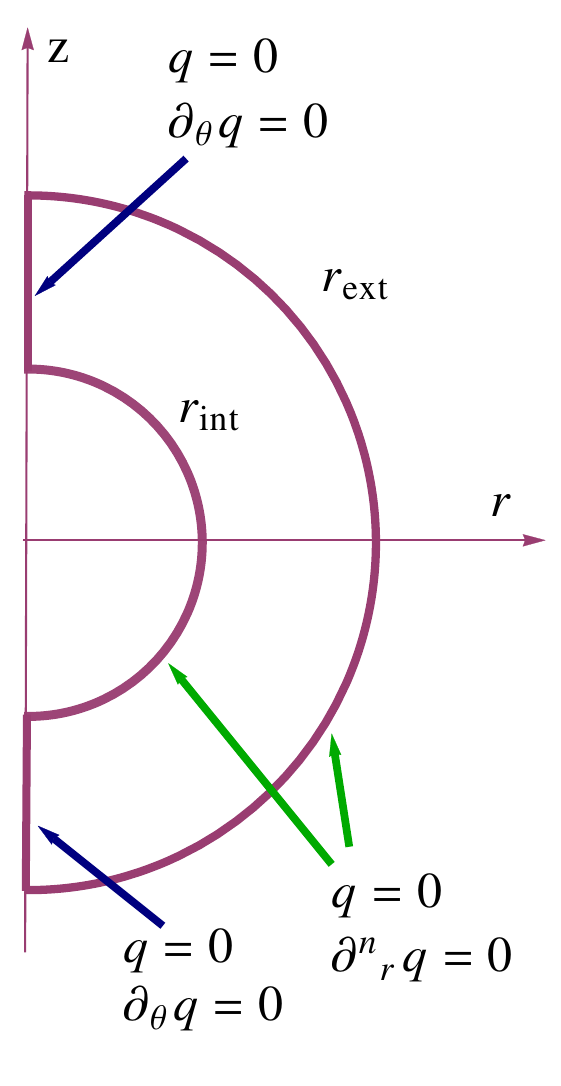}
 \caption{The boundary conditions. The thick lines indicate the loci where boundary conditions 
 on the function $q$ must be imposed. On the z-axis the conditions \eqref{bound_cond_theta} 
 related to the axisymmetry of our construction must be satisfied while on the boundaries 
 of the gluing annulus the conditions \eqref{bound_cond_r} must be implemented.}
 \label{fig:bound_cond}
\end{figure}

The function $q(r,\theta)$ in the gluing region will be determined by Einstein's equations. 
In addition to the boundary conditions \eqref{bound_cond_theta} on the $\textrm{z}$-axis, 
smoothness requires that $q$ and all its radial derivatives vanish at the boundaries 
of the gluing annulus: 
\begin{equation}
\label{bound_cond_r}
 q = 0 \quad \mathrm{and} \quad \frac{\partial^n \! q}{\partial r^n} = 0 \qquad \mathrm{at} \quad
 r = r_{\mathrm{int}}, r_{\mathrm{ext}},
\end{equation}
for all $n \in \mathbb{N}$. 
The boundary conditions that $q$ must satisfy are summarised in Fig.~\ref{fig:bound_cond}.

\subsection{Mathematical fomulation}
\label{sec:math_formulation}

Having set up our gluing scheme in the previous sections, we now move on to Einstein's 
equations. 
On the initial slice these reduce to the momentum and Hamiltonian 
constraints. The former is identically satisfied as our data are time-symmetric, so we are 
left only with the Hamiltonian constraint, which in the time-symmetric case reduces to the 
vanishing of the Ricci scalar of the Brill wave metric \eqref{Brill_metric}, 
i.e.
\begin{equation*}
 R({g}_{\textrm{\tiny Brill}}) = 0.
\end{equation*}
Expanding the Ricci scalar in the above expression, the Hamiltonian constraint results in 
an inhomogeneous Poisson equation of the form
\begin{eqnarray}
\label{Poisson_eq}
 {}^{(2)} \Delta q &=& -4\, \frac{{}^{(3)}\Delta \psi}{\psi} \quad :\Leftrightarrow \nonumber\\
 \frac{\partial^2 q}{\partial r^2} + \frac{1}{r^2} \frac{\partial^2 q}{\partial \theta^2} + \frac{1}{r} 
 \frac{\partial q}{\partial r} &=& -\frac{4}{\psi} \left( \frac{\partial^2 \psi}{\partial r^2} + \frac{1}{r^2} \frac{\partial^2 \psi}{\partial \theta^2} +
  \frac{2}{r} \frac{\partial \psi}{\partial r} + \frac{\cot\theta}{r^2} \frac{\partial \psi}{\partial \theta} \right) =: f.
\end{eqnarray}
According to our construction in Sec.~\ref{sec:recipe}, the right-hand side of the above 
elliptic equation is specified by the form of $\psi$ that is defined by \eqref{conf_factor} 
and \eqref{beta_function}. Since this is fixed \textit{a priori}, we will consider the 
right-hand side of \eqref{Poisson_eq} as an inhomogeneity and denote it by $f$. It should 
be noted that \eqref{Poisson_eq} reduces to a homogeneous Poisson equation outside the 
gluing annulus as the constancy of $\beta$ enforces $f$ to vanish there. Summarising, 
our goal in the following will be to numerically solve the second-order linear PDE \eqref{Poisson_eq} 
for $q(r,\theta)$ subject to the boundary conditions \eqref{bound_cond_theta} and \eqref{bound_cond_r}.

\subsection{Integrability condition}
\label{sec:integrability}

At first sight it might seem that the choice of the two mass parameters $m$ and $M$ 
in the conformal factor \eqref{conf_factor} is unconstrained. If this were true then 
nothing would prevent us from gluing a Minkowskian end to the Brill-Lindquist data 
in the interior! This would obviously violate the positive mass theorem \cite{Sch&Yau}. 
In fact Einstein's equations constrain the choice of the masses. One way to see this 
is by employing the machinery developed by Brill \cite{Brill:1959} in order to prove 
that the ADM mass of time-symmetric, axisymmetric, vacuum gravitational waves is positive 
definite. It turns out that in our setting this result can be used as a condition to 
determine the relation between the masses involved in our construction. 

Following Brill's arguments in \cite{Brill:1959}, we repeat here his original derivation 
adjusted to the details of our construction. Our starting point is the Poisson equation 
\eqref{Poisson_eq} expressed in cylindrical coordinates $(\rho, \phi, z)$:
\begin{equation*}
 \frac{\partial^2 q}{\partial \rho^2} + \frac{\partial^2 q}{\partial z^2} +
 \frac{4}{\psi} \left( \frac{\partial^2 \psi}{\partial \rho^2} + \frac{\partial^2 \psi}{\partial z^2} +
  \frac{1}{\rho} \frac{\partial \psi}{\partial \rho} \right) = 0, 
\end{equation*}
which when expressed in terms of the three-dimensional flat Laplace operator 
$\nabla^2 = \rho^{-1} \partial_\rho + \partial^2_\rho + \rho^{-2} \partial^2_\phi + \partial^2_z$
in cylindrical coordinates takes the form
\begin{equation*}
 4\, \frac{\nabla^2 \psi}{\psi} + \nabla^2 q - \frac{1}{\rho} \frac{\partial q}{\partial \rho} = 0. 
\end{equation*}
Integrating over the interior $V$ of a large sphere $\Sigma$ of radius $R$ centred at 
the origin, one gets
\begin{equation}
 \label{Poisson_cyl}
  4\int_V \left[ \nabla \cdot \left( \frac{\nabla \psi}{\psi} \right) + \left( \frac{\nabla \psi}{\psi} \right)^2 \right] dV + 
  \int_V \nabla^2 q\, dV - \int_V \frac{1}{\rho} \frac{\partial q}{\partial \rho}\, dV = 0, 
\end{equation}
where the gradient and the divergence in cylindrical coordinates read
$\nabla = (\partial_\rho, \rho^{-1} \partial_\phi, \partial_z)$ 
and $\nabla \cdot = (\rho^{-1} + \partial_\rho, \rho^{-1} \partial_\phi, \partial_z) \cdot \,$, respectively.
The integration of the last term reads
\begin{equation*}
 \int_V \frac{1}{\rho} \frac{\partial q}{\partial \rho}\, \rho\, d\rho\, d\phi\, dz = 
 2\, \pi \int_{-R}^R \left[ q\left(\sqrt{R^2 - z^2}, z\right) - q(0, z) \right] dz = 
 - 2\, \pi \int^\pi_0 q(R, \theta)\, R\, \sin\theta\, d\theta, 
\end{equation*}
where in the last step we used the first of the boundary conditions \eqref{bound_cond_theta} 
and expressed the remaining term in spherical coordinates. In the rest of the proof, the 
first two integrals in \eqref{Poisson_cyl} will also be expressed in spherical coordinates. 
Inserting the result of the above integration into \eqref{Poisson_cyl} and re-expressing 
the first and third term through the divergence theorem, one arrives at
\begin{equation}
  8\, \pi \int^\pi_0 \left. \frac{1}{\psi}\frac{\partial \psi}{\partial r}\right|_{r=R} R^2 \sin\theta\, d\theta +
  4\int_V   \left( \frac{\nabla \psi}{\psi} \right)^2  dV + 
  2\, \pi \int^\pi_0 \left[ \left. \frac{\partial q}{\partial r}\right|_{r=R} R^2 + q(R, \theta)\, R \right] \sin\theta\, d\theta = 0. 
  \label{Poisson2}
\end{equation}
In the limit $R \rightarrow \infty$ the last term of the above expression vanishes because
$q = 0$ for $r > r_\textrm{ext}$. In addition, 
according to \eqref{conf_factor}, the conformal factor in the limit $R \rightarrow \infty$ 
behaves like $1 + \frac{M}{2\, R}$; thus the first term of the expession above reads
\begin{equation*}
 8\, \pi \int^\pi_0 \left. \frac{1}{\psi}\frac{\partial \psi}{\partial r}\right|_{r=R} R^2 \sin\theta\, d\theta = 
 8\, \pi \int^\pi_0 \frac{-\frac{M}{2\, R^2}}{1 + \frac{M}{2\, R}} R^2 \sin\theta\, d\theta \overset{R \rightarrow \infty}{=}
 - 8\, \pi\, M.
\end{equation*}
Taking into account the last two results, \eqref{Poisson2} in the limit $R \rightarrow \infty$ 
reduces to 
\begin{equation*}
  - 2\, \pi\, M + \int_V   \left( \frac{\nabla \psi}{\psi} \right)^2 dV  = 0. 
\end{equation*}  
Finally, expanding the integrand and integrating over $\phi$ one arrives at Brill's original 
expression for the ADM mass,
\begin{equation}
 \label{integr_cond}
  M = \int^\pi_0 \int^\infty_0 \left[ \left( \frac{1}{\psi}\frac{\partial \psi}{\partial r} \right)^2 + 
  \left( \frac{1}{r\, \psi}\frac{\partial \psi}{\partial \theta} \right)^2 \right] r^2 \sin\theta\, dr\, d\theta,
\end{equation}
which is obviously positive definite. It is interesting that this expression for the ADM 
mass only depends on the conformal factor. Recall that the ADM mass $M$ of our construction 
appears in the definition of the conformal factor \eqref{conf_factor} and consequently is 
also present in the integrand above. Based on this observation one can use \eqref{integr_cond} 
as an integrability condition for the ADM mass, namely the integral on the right-hand side 
of \eqref{integr_cond} for a specific choice of $M$ must return the same value for the ADM 
mass.


\section{Numerical implementation of the gluing construction}
\label{sec:numer_implement}

In this section our numerical implementation of the gluing construction described 
in the previous section and some first numerical results are presented.

\subsection{Setting up the numerical scheme}
\label{sec:numer_scheme}

We choose to solve the Poisson equation \eqref{Poisson_eq} numerically using pseudo-spectral 
methods. Accordingly, the unknown function $q(r, \theta)$ is approximated by a truncated 
series of suitable specific polynomials. We choose to expand the $r$-dependence of $q$ in 
Chebyshev polynomials $T_k$ and the $\theta$-dependence in Fourier-cosine 
series for reasons (in addition to the ones presented in \cite{Bo&Gou&Ma:1999}) 
that will soon become apparent. 

Our two-dimensional physical domain is given by $(r, \theta) \in [r_{\mathrm{int}}, r_{\mathrm{ext}}] \times [0, \pi]$. 
While the range of the angular coordinate $\theta$ is in accordance with the expansion in Fourier-cosine series, the range of the radial coordinate $r$ is not, 
as the Chebyshev polynomials are defined on the interval $[-1, 1]$. 
In order to map the original $r$-domain to $[-1, 1]$, we use the mapping 
\begin{equation*}
 x \mapsto r(x) := \frac{1}{2}(r_{\mathrm{ext}} - r_{\mathrm{int}})\, x + \frac{1}{2}(r_{\mathrm{ext}} + r_{\mathrm{int}}), 
\end{equation*}
where $x$ takes values in the interval $x \in [-1, 1]$. Therefore, from now on, we have to 
think of the expressions \eqref{conf_factor}, \eqref{beta_function} and \eqref{Poisson_eq} 
as expressed in terms of this new linearly transformed radial coordinate $x$. Therefore, 
in the following our two-dimensional computational domain will be $D = [-1, 1] \times [0, \pi]$. 
A finite representation of $D$ is obtained by the introduction of equidistant collocation 
points in the $\theta$-direction and of non-equidistant Gauss-Lobatto collocation points 
in the radial direction, namely
\begin{equation*}
 \theta_i = \frac{i\, \pi}{L} \qquad \mathrm{and} \qquad x_j =  -\cos\left(\frac{j\, \pi}{K}\right) 
 \qquad \mathrm{with} \qquad i = 0, \ldots, L \quad \mathrm{and} \quad j = 0, \ldots, K,
\end{equation*}
where $K$ and $L$ is the number of collocation points along the radial and $\theta$-direction, 
respectively.

Let us now turn to the boundary conditions \eqref{bound_cond_theta} and \eqref{bound_cond_r}. 
In fact this is by far the most involved part of our numerical implementation. In order 
to satisfy \eqref{bound_cond_r} we make the following ansatz:
\begin{equation}
 \label{ansatz}
  q(x, \theta) = B(x)\, \hat{q}(x, \theta),
\end{equation}
where $\hat{q}$ is an arbitrary function of its arguments and $B(x)$ is a function of 
``bump'' character on the gluing annulus, i.e. $B(x)$ and all its $x$-derivatives vanish 
on the boundaries of the gluing annulus. An example of a ``bump'' function with the above 
properties looks like 
\begin{equation}
\label{bump_function}
 B(x) = \mathrm{sech} \left(\frac{b_1}{x - 1} + \frac{b_2}{x + 1} \right),
\end{equation}
where $b_1, b_2$ are constants. 
The convergence of our numerical solutions crucially depends on the choice of these constants. 
It has been observed that the convergence properties of the produced numerical solutions are 
optimal when the constants $b_1, b_2$ take values $b_1, b_2 < 1$. In the following the choice 
$b_1 = b_2 = 10^{-2}$ will always be used. The second boundary condition in \eqref{bound_cond_theta} 
is satisfied if one expands the newly introduced function $\hat{q}(x, \theta)$ in the way 
described in the first paragraph of this section, namely
\begin{equation}
\label{q_expansion}
 \hat{q}(x, \theta) = \sum^K_{k=0}{\sum^{L}_{l=0}{a_{kl}\, T_k(x)\, \cos(l\, \theta)}}, 
\end{equation}
where $K, L$ are as above and the constants $a_{kl}$ are the expansion coefficients of our 
series. In order to satisfy the remaining boundary condition, i.e. the first of \eqref{bound_cond_theta}, 
one can use the freedom inherent in the choice of the gluing function \eqref{beta_function}. 
Recall that the gluing function, apart from the specific conditions that it has to satisfy 
on the boundaries of the gluing annulus, can be freely specified otherwise. A possible ansatz 
is
\begin{equation}
 \label{ansatz_beta}
  \beta(x, \theta) = \alpha(x) + \hat{\alpha}(x) B(x) \sin^2\theta,
\end{equation}
where 
\begin{equation*}
   \alpha(x) = \frac{1}{2} \left(1 + \tanh \left(\frac{1}{x - 1} + \frac{1}{x + 1}\right)\right),
\end{equation*} 
$B(x)$ is given by \eqref{bump_function} and $\hat{\alpha}(x)$ 
is a so far arbitrary function that we choose in order to enforce the condition $q = 0$ on the 
z-axis. Notice that the function $\alpha(x)$ takes the values $1$ and $0$ on the internal $x=-1$ 
and external $x=1$ boundary of the gluing annulus, respectively, and all its spatial derivatives 
vanish there; thus, it satisfies all the criteria of \eqref{beta_function}. The inclusion of the 
``bump'' function $B(x)$ in the ansatz \eqref{ansatz_beta} guarantees that, independently of the 
choice of $\hat{\alpha}(x)$, the second term in \eqref{ansatz_beta} and all its derivatives vanish 
identically on the boundaries. Therefore, the form of $\hat{\alpha}(x)$ influences the shape of 
$\beta(x, \theta)$ only in the interior of the gluing annulus. 
(It is noteworthy that with a $\theta$-independent ansatz, e.g. of the form $\beta(x) = \alpha(x) + \hat{\alpha}(x) B(x)$, 
it was not possible to satisfy the first boundary condition in \eqref{bound_cond_theta} and at the same time have a convergent numerical solution.)
Now, as the roots of the map 
\begin{equation}
 \label{alphahat_map}
  \hat \alpha(x) \mapsto q(x, \theta \in \{0,\pi\})
\end{equation}
are $(K+1)$-dimensional vectors (recall $K$ refers to the number of radial collocation points),
we have to use a multidimensional secant (quasi-Newton) method to find them. (We chose to use 
a secant instead of a Newton method as the former is computationally less costly and faster.) 
The most effective and efficient method of this kind has proven \cite{numer_recipes2007} to be 
\emph{Broyden's method} \cite{Broyden1965}. Given an initial guess for $\hat \alpha(x)$, Broyden's 
method tries to find iteratively the form of $\hat \alpha(x)$ that leads to a solution of \eqref{Poisson_eq} 
satisfying the first boundary condition in \eqref{bound_cond_theta} to a given accuracy 
(here to the order of $\sim 10^{-14}$). In the following the roots of \eqref{alphahat_map} 
will be computed numerically using the implementation of Broyden's method in the optimize 
sub-package of Python's SciPy library.

Summarising, by assuming that $q$ in \eqref{ansatz} is a multiple of a ``bump'' function 
$B$, the vanishing of $q$ and all its  $x$-derivatives at $x = \pm 1$ is guaranteed. The 
expansion of $\hat{q}$ as a Fourier-cosine series sets $\partial_\theta \hat{q}$ to zero 
on the $z$-axis; consequently $\partial_\theta q$ also vanishes there as $\partial_\theta q = B\, \partial_\theta \hat{q}$. 
Finally, an appropriate choice of the function $\hat{\alpha}(x)$ in the ansatz \eqref{ansatz_beta} 
can make $q$ vanish on the $z$-axis. 

So far we have assumed that the mass parameter $M$ appearing in the conformal factor \eqref{conf_factor} 
is given. However, this parameter has to agree with the integral expression \eqref{integr_cond} 
for the ADM mass, which contains the conformal factor---hence $M$ is only given implicitly.
We start by choosing an initial value for $M$ and solve for $q$ and $\hat\alpha$ using the 
iterative procedure described above. Knowing $\hat \alpha$ and thus the gluing function $\beta$, 
we can compute the conformal factor \eqref{conf_factor} and evaluate the value $M_I$ of the 
integral for the ADM mass \eqref{integr_cond}. Then we vary $M$ until a value satisfying 
$M=M_I$ is found, repeating the above procedure at each step. This will be illustrated in 
Sec.~\ref{sec:reduction}.

The code has been written from scratch in Python.

\subsection{Testing the code with an exact solution}
\label{sec:exact_sol}

Before we start using our code to study numerically the Poisson equation \eqref{Poisson_eq}, 
we will carry out---as one should always do---some numerical tests to check the performance 
of our code. For this a family of exact solutions will be used. The exact solutions will be 
computed in the following way. First, we choose a $q$ and compute analytically the outcome 
of the left-hand side of \eqref{Poisson_eq}, then we equate the resulting expession with the inhomogeneity 
$f$. Now, having at hand the expression for $f$, one can solve numerically \eqref{Poisson_eq} 
for $q$ and compare the outcome with the exact expression of $q$ chosen originally. This 
procedure will give us hints about the accuracy and the convergence properties of the code. 
\begin{figure}[htb]
 \centering
  \subfigure[]{
   \includegraphics[scale = 0.25]{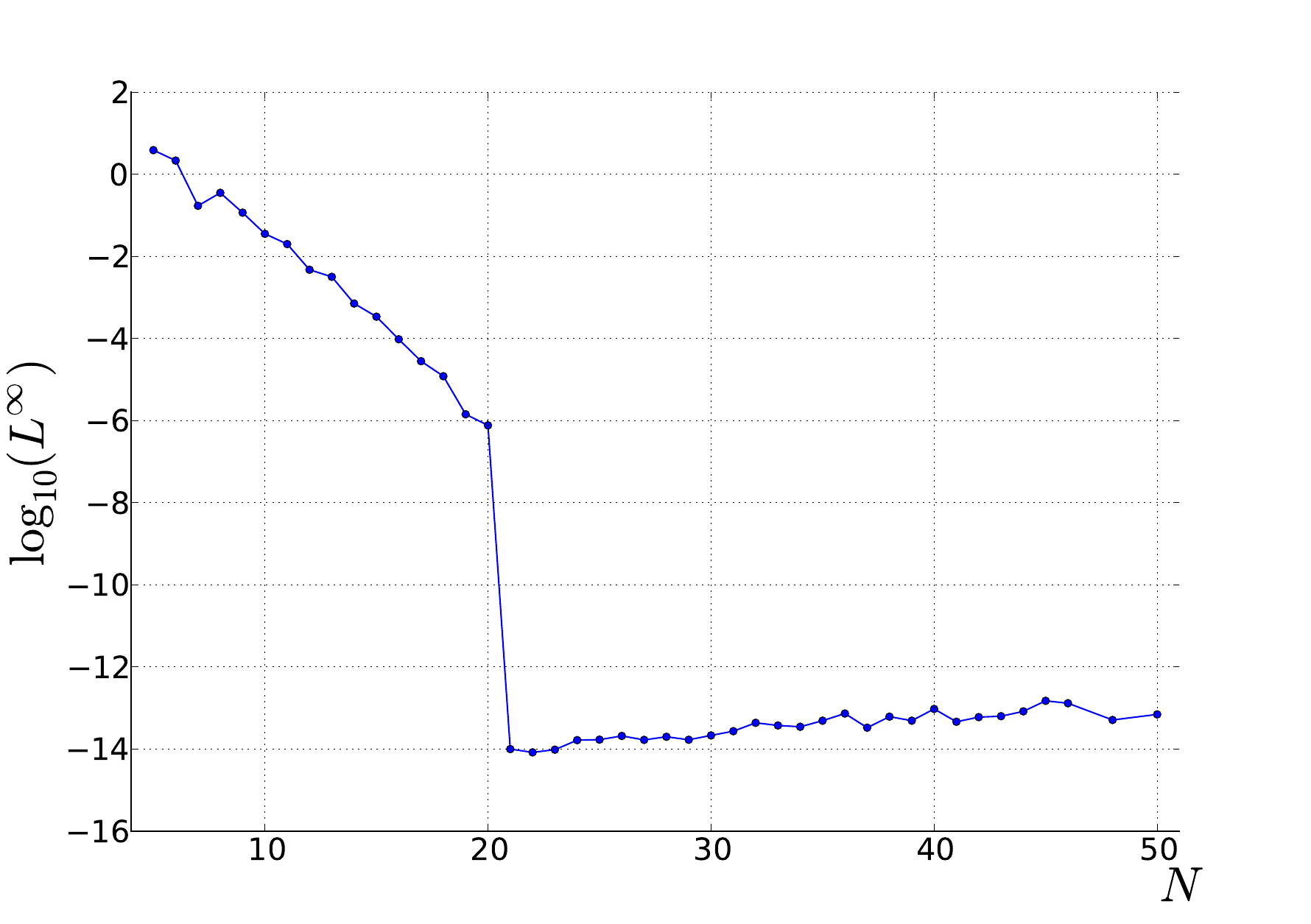}
   \label{fig:exact_a}
  }
  \subfigure[]{
   \includegraphics[scale = 0.25]{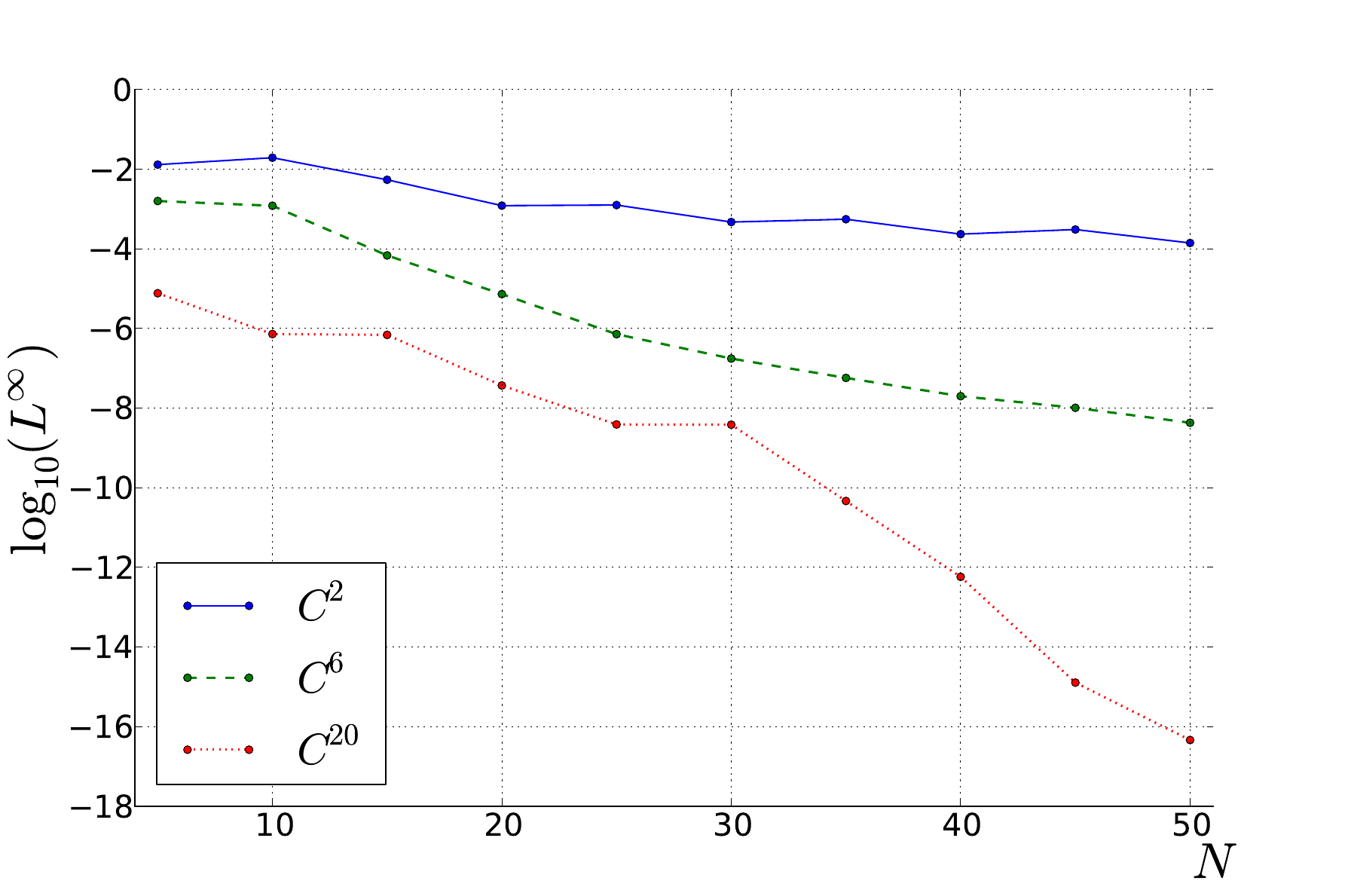}  
   \label{fig:exact_b}
  }
 \caption{Convergence against the exact solutions \eqref{q_exact}. Shown is the $L^\infty$ 
 norm of the error as a function of the number of grid points $N$. (a) The exact solution 
 $q$ is smooth, i.e. infinitely differentiable $\mathcal{C}^\infty$. The characteristic 
 ``step'' behaviour of polynomial functions is clearly visible. (b) The exact solution $q$ 
 is of finite differentiability. As expected the higher the differentiability of the solution, 
 the faster the convergence.
 }
 \label{fig:exact}
\end{figure}

As exact solutions we will use the following family of functions,
\begin{equation}
\label{q_exact}
 q(x, \theta) = x^{\kappa/3} \left(x - 1\right)^{10} \left(x + 1\right)^{10} \sin(6\, \theta) B(x) \sin\theta, 
\end{equation}
where $\kappa$ is a non-negative integer and $B$ is the ``bump'' function \eqref{bump_function}. 
The main reason for choosing the above family of solutions is that it allows us to control, 
through the choice of $\kappa$, the differentiability, and consequently the smoothness, at 
$x = 0$. Obviously, if $\kappa$ is zero or a multiple of three, then the function $\hat q$ 
corresponding to \eqref{q_exact} is a polynomial and thus infinitely differentiable $\mathcal{C}^\infty$. 
For any other value of $\kappa$, \eqref{q_exact} is finitely differentiable $\mathcal{C}^j$. In 
the following, we will assume that $\kappa$ takes the values 
$\kappa = 0, 7, 19, 61$
and as a consequence the solution \eqref{q_exact} will be 
$\mathcal{C}^\infty, \mathcal{C}^2, \mathcal{C}^6, \mathcal{C}^{20}$ 
at $x = 0$, respectively. 
Our goal of doing all this is not only to show that the numerical solutions converge to 
the exact ones, but also to observe the expected relation, see e.g.~\cite{Trefethen:2000}, 
between the convergence of the numerical solutions and the smoothness of the exact solution, 
i.e. the smoother the solution \eqref{q_exact}, the faster the convergence of the code.

Our findings are presented in Fig.~\ref{fig:exact}. Both graphs therein depict the $\log_{10}$ 
of the absolute value of the maximum error (in other worlds the $L^\infty$ norm) between 
the numerical and the corresponding exact solution for different numbers of grid points 
$N$, where here 
we have chosen $K = L =: N$. 
Fig.~\ref{fig:exact_a} illustrates the case of smooth functions ($\kappa=0$). 
Here one observes the typical ``step'' behaviour of the convergence plots corresponding 
to polynomial functions \cite{Trefethen:2000}; this is because the Chebyshev polynomials 
form a complete basis for the polynomials, so that \eqref{q_exact} is represented \textit{exactly} 
for $N>20$ (the error settles down to numerical roundoff 
$\sim 10^{-14}$). 
On the other hand, Fig.~\ref{fig:exact_b} shows the case of finitely differentiable functions. It 
can be easily seen that in all the cases considered the numerical solutions converge to 
the exact ones, but with different speed. A detailed inspection of the individual plots 
shows that, as expected, the speed of convergence is faster the smoother is our solution 
\cite{Trefethen:2000}.

\subsection{Numerical realisation of the gluing construction}
\label{sec:numer_gluing_constr}

\subsubsection{Results}
\label{sec:results}

The results of the previous section constitute strong evidence that our code can reproduce 
successfully the exact solutions \eqref{q_exact}, and its convergence behaviour is as 
expected. Thus, we are confident enough to proceed further in the numerical study of 
the gluing construction and look for general solutions of \eqref{Poisson_eq}.
\begin{figure}[!htb]
 \centering\hspace{-8mm}
  \subfigure[]{
   \includegraphics[scale = 0.16]{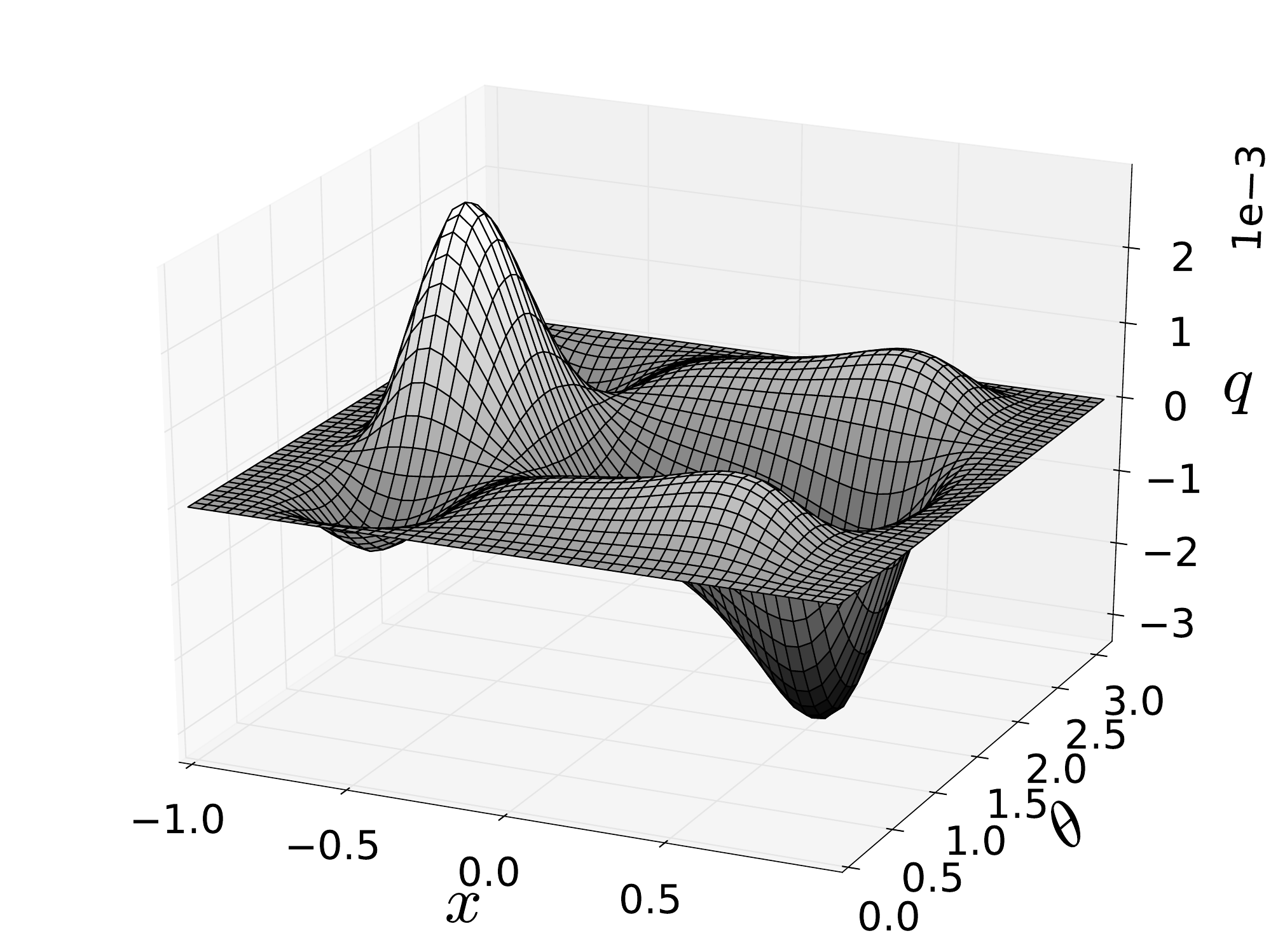} 
   \label{fig:num_sol_a}
  }\hspace{-3.5mm}
  \subfigure[]{
   \includegraphics[scale = 0.16]{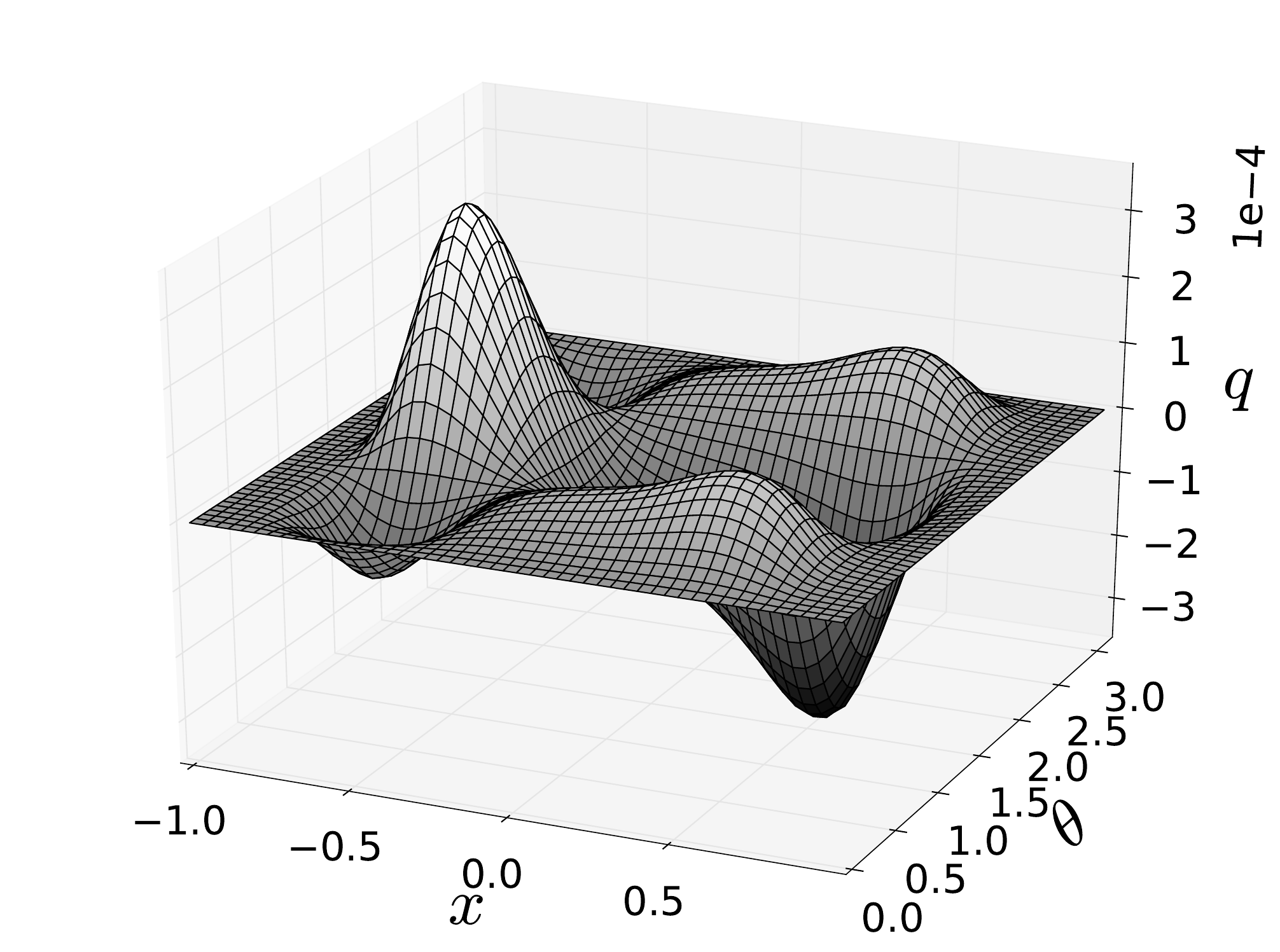} 
   \label{fig:num_sol_b}
  }\hspace{-3.5mm}
  \subfigure[]{
   \includegraphics[scale = 0.16]{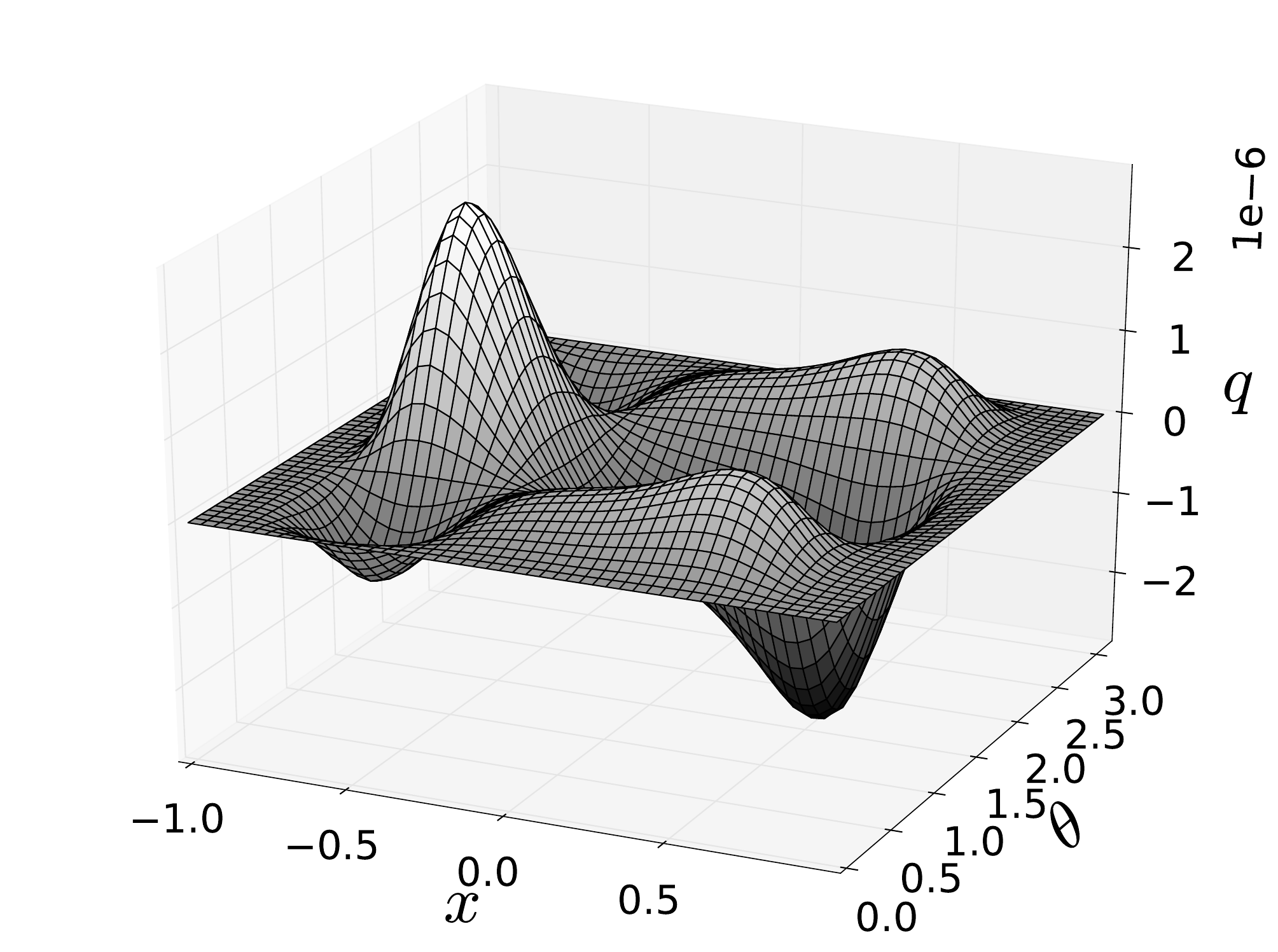} 
   \label{fig:num_sol_c}
  }
 \caption{Numerical solutions of \eqref{Poisson_eq}. The function  $q$ is plotted on 
 the grid $D = [-1,1]\times[0,\pi]$ for the choice $m = 2$, $d = 10$, $r_{\mathrm{ext}} = 
 2\, r_{\mathrm{int}}$, and gluing radii (a) $r_{\mathrm{int}} = 50$,  (b) $r_{\mathrm{int}} = 100$, 
 (c) $r_{\mathrm{int}} = 500$. The corresponding ADM masses are (a) $M = 4.001084$, 
 (b) $M = 4.00002725$ and (c) $M = 4.0000000095$, see Fig.~\ref{fig:ADM_increase}. 
 Notice the order of magnitude of the solutions; as expected, it decreases with 
 the distance of the gluing annulus from the origin. The numerical resolution used 
 here is $K=L=25$ collocation points.
 }
 \label{fig:num_solution}
\end{figure}

In order to do so, one has first to choose appropriately the free parameters entering 
the definition of the conformal factor \eqref{conf_factor} and then to compute the 
inhomogeneity $f$ by evaluating the right-hand side of \eqref{Poisson_eq}. Recall that 
according to its definition, the conformal factor depends on the mass $m$ of the individual 
Brill-Lindquist black holes, their mutual distance $d$, the mass $M$ of the exterior 
Schwarzschild region, the location of the gluing annulus $r_{\mathrm{int}}, r_{\mathrm{ext}}$, 
and the form of the gluing function \eqref{beta_function}. 
In the following, the ansatz \eqref{ansatz_beta} will be used for the gluing function and the form of $\hat \alpha(x)$ entering its definition will be computed in accordance 
with the discussion of Sec.~\ref{sec:numer_scheme}. 
Except for a couple of conditions 
that constrain their choice, the above parameters can be freely chosen. The first condition 
follows from the fact that the gluing annulus has to be placed away from any horizons 
of the Brill-Lindquist data; for this the inequality $r_{\mathrm{int}} > d/2 + r_\mathrm{hor}$ 
must always be satisfied---see Sec.~\ref{sec:basics} for the details. The second condition 
constrains the relation of the masses $m$ and $M$, as discussed in Sec.~\ref{sec:integrability}.

Fig.~\ref{fig:num_solution} shows several numerical solutions of the system 
\eqref{Poisson_eq}, \eqref{bound_cond_theta}, \eqref{bound_cond_r} for the 
following choice of the free parameters: $m = 2$, $d = 10$, $r_{\mathrm{ext}} = 2\, r_{\mathrm{int}}$, 
and the ADM mass $M$ has been chosen such that the integrability condition 
\eqref{integr_cond} is satisfied (see Fig.~\ref{fig:ADM_increase}). Starting 
from Fig.~\ref{fig:num_sol_a}, the distance of the gluing annulus from the 
origin has been gradually increased from $r_{\mathrm{int}} = 50$ to $r_{\mathrm{int}} = 500$. 
As expected, the further away one places the gluing annulus, the smaller 
the numerically computed values of $q$ become. This behaviour follows naturally 
from the fact that the Brill-Lindquist data \eqref{B-L_metric} look more 
and more like Schwarzschild data the further away one goes from the origin; 
consequently, the Brill wave---essentially the function $q$---does not have 
to do ``a lot of work'' to glue the two sets of data together. Similar behaviour 
is observed when the distance of the annulus from the origin is kept fixed 
but its width is gradually increased. Now, the magnitude of $q$ gradually 
decreases as it has ``more and more space'' to perform the gluing between 
the two data sets. 

The results of Fig.~\ref{fig:num_solution} are the first evidence that the 
gluing constructions proposed in \cite{Corv:2000, Giu&Hol:2005} can be realised 
numerically. Whereas the analysis of \cite{Corv:2000, Giu&Hol:2005} applies 
only to the case when the gluing annulus is placed at large distances, our 
numerical findings here demonstrate that these results can be extended to 
smaller gluing radii.

At this point, it is worth checking what happens in the case that the distance between 
the two black holes is taken to be $d=0$ so that there is only a single black hole 
of mass $2m$ in the centre. One would expect that as long as the condition $M = 2m$ 
is satisfied, the function $q$ must vanish; for in this setting the Brill-Lindquist 
data \eqref{B-L_metric} are already in Schwarzschild form. It turns out that our code 
correctly reproduces the trivial solution for arbitrary position of the gluing annulus.

\subsubsection{Convergence analysis}
\label{sec:con_analysis}

Let us turn now to the convergence analysis of our numerical solutions. In contrast to 
Sec.~\ref{sec:exact_sol}, here we do not have an exact solution to compare our numerical 
findings with. Thus, we have to follow a different approach to check the convergence of our 
numerical solutions. The usual way to proceed in such a situation is to study the decay 
of the expansion coefficients $a_{kl}$ in \eqref{q_expansion}, see \cite{Boyd:2001}. The 
expansion coefficients $a_{kl}$ must gradually decay to zero for increasingly large indices
in order for the series expansion \eqref{q_expansion} to converge. 
Once $q$ has been computed numerically, the expansion 
coefficients can be readily evaluated by inverting \eqref{q_expansion}. 
\begin{figure}[htb]
 \centering
  \subfigure[]{
   \includegraphics[scale = 0.26]{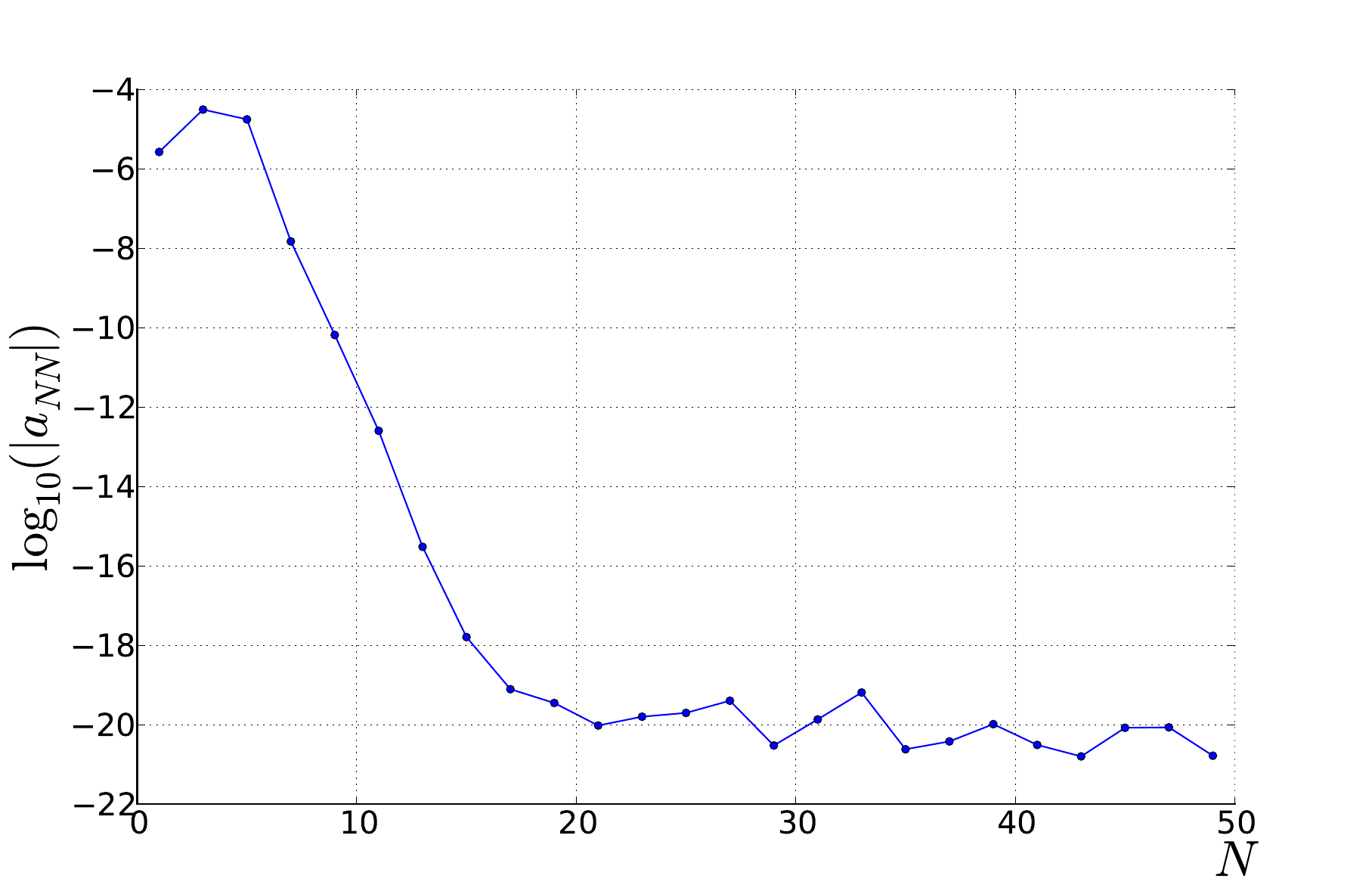}
   \label{fig:num_decay}
  }
  \subfigure[]{
   \includegraphics[scale = 0.26]{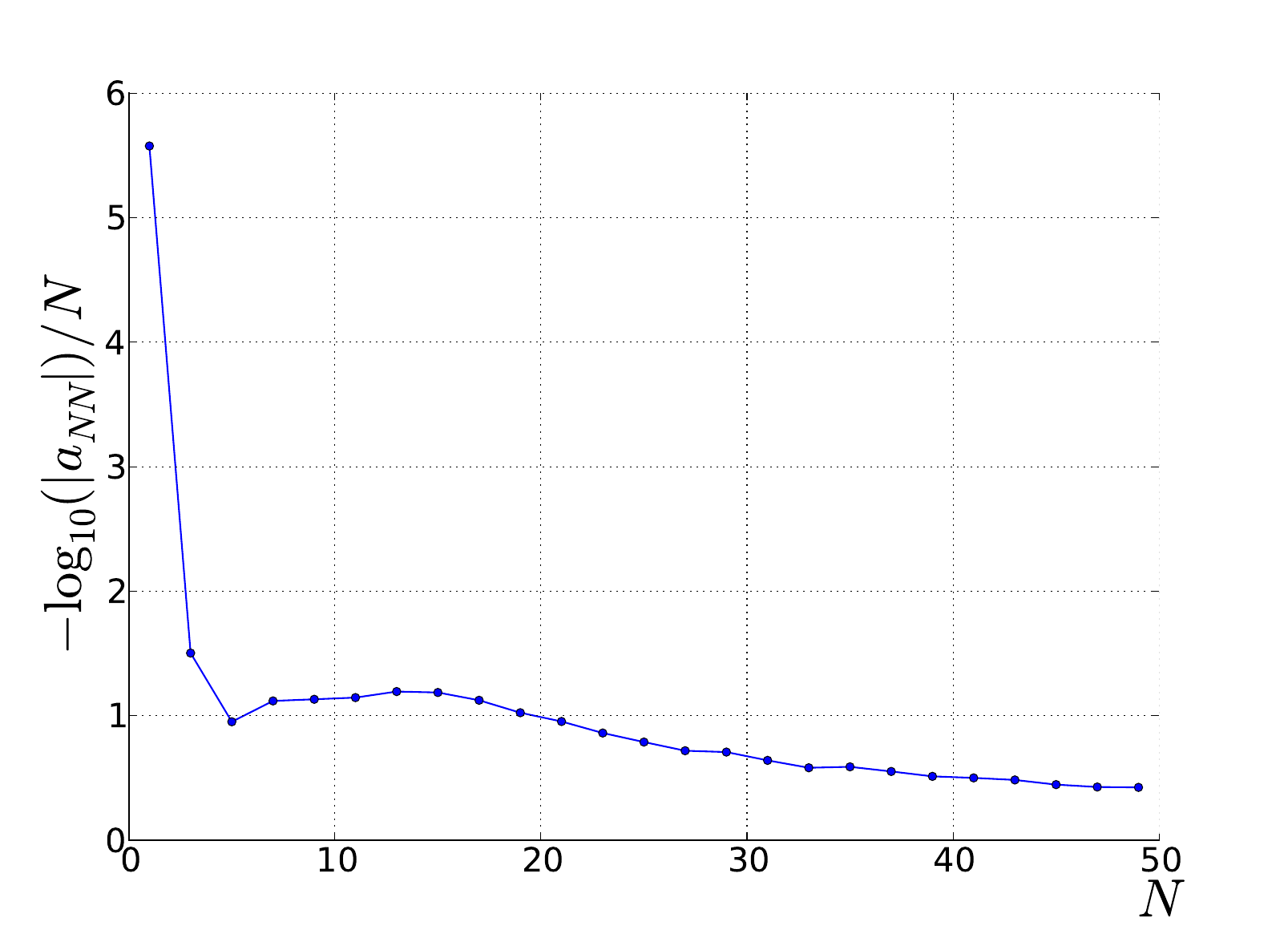}  
   \label{fig:num_exp_conv}
  }
 \caption{Exponential decay of the expansion coefficients with the number of grid points 
 $N$. (a) The diagonal expansion coefficients for $N < 20$
 fall off in an almost linear fashion in 
 this logarithmic plot, indicating exponential decay. (b) Testing for exponential decay. 
 The ratio $-\log_{10}(|a_{NN}|)/N$ must asymptote to a non-negative number in the limit 
 $N \rightarrow \infty$.}
 \label{fig:num_convergence}
\end{figure} 
Fig.~\ref{fig:num_convergence} depicts the results of our convergence analysis for the 
numerical solution of Fig.~\ref{fig:num_sol_b}. Because of the two-dimensional nature 
of the series expansion \eqref{q_expansion}, one has to choose along which direction 
to study $a_{kl}$.  We chose here to study the convergence behaviour of the diagonal 
expansion coefficients $a_{NN}$ as they provide a good indication of the overall decay 
of $a_{kl}$. The fall-off behaviour of $|a_{NN}|$ is depicted in Fig.~\ref{fig:num_decay} 
on a logarithmic scale; the observed approximately linear behaviour for $N < 20$ suggests 
an exponential decay to the roundoff plateau. To make this statement more quantitative, 
one has to study the ratio $-\log_{10}(|a_{NN}|)/N$ in the limit $N \rightarrow \infty$. 
Therefore, following \cite{Boyd:2001}, if the limit 
\begin{equation*}
 \lim_{N \to \infty}\left(\frac{-\log_{10}(|a_{NN}|)}{N}\right) \geq 0
\end{equation*}
is a non-negative number then the expansion coefficients converge to zero exponentially. 
In Fig.~\ref{fig:num_exp_conv} one clearly sees a tendency of the ratio $-\log_{10}(|a_{NN}|)/N$ 
to asymptote to a small positive number, which is a strong indication of exponential decay. 

\begin{figure}[htb]
 \centering
 \includegraphics[scale = 0.25]{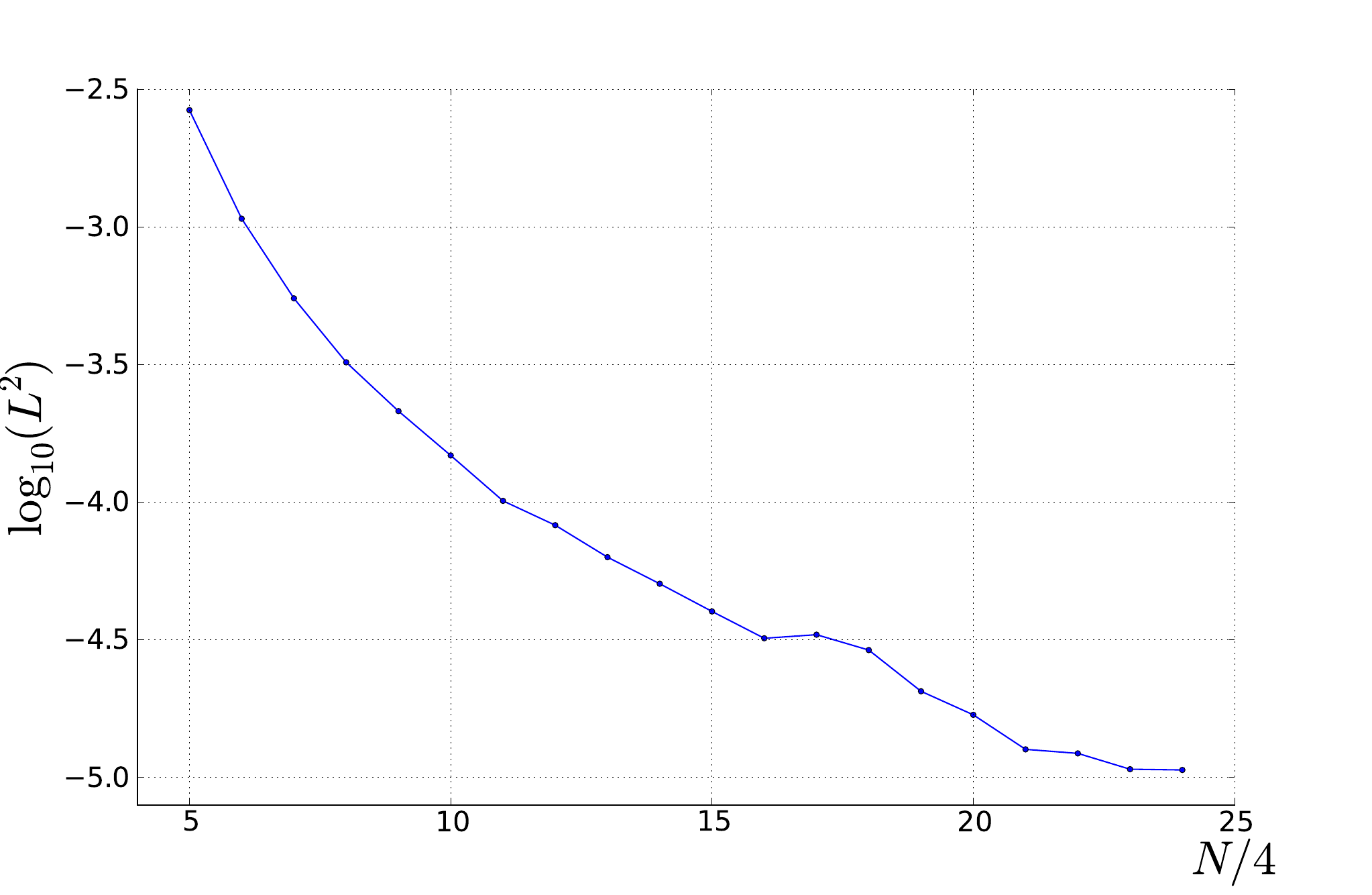}
 \caption{
 Convergence with respect to the numerical solution of highest resolution. 
 The logarithm of the error falls off roughly linearly with the number of grid points 
 $N/4$ along the angular direction.
 }
 \label{fig:num_high_resolution}
\end{figure}

We will conclude the present section by presenting another indication that the numerical 
solutions produced in Sec.~\ref{sec:results} converge exponentially. In Fig.~\ref{fig:num_high_resolution}, 
on a rectangular $N \times N/4$ grid (i.e.~$N$ grid points along the radial and $N/4$ 
along the angular direction), we compare numerical solutions of different resolutions 
to the one with the highest resolution for the solution of Fig.~\ref{fig:num_sol_b}. 
Specifically, the numerical values of $q$ for each resolution are interpolated onto 
the same grid and compared with the solution of highest resolution there (here an 
$100\times 25$ grid). Finally, the $L^2$-norm of the absolute value of the error for 
each resolution has been plotted on a logarithmic scale, see Fig.~\ref{fig:num_high_resolution}. 
The curve falls off in an approximately linear fashion.

\subsection{Behaviour of the ADM mass}
\label{sec:reduction}

We will now investigate the dependence of the ADM mass on the details of the gluing 
construction. Namely, we examine if it is possible to choose the free parameters entering 
the definition of the conformal factor \eqref{conf_factor} in such a way that the ADM 
mass $M$ can take values different from the sum of the two Brill-Lindquist black holes, 
i.e. $M \neq 2\, m$. The case $0 < M < 2\, m$ corresponds to a reduction of the ADM 
mass, while the case $M > 2\, m$ to an increase. In other words, we explore the possibility 
of gluing together the spacetimes \eqref{B-L_metric} and \eqref{Schw_metric} under 
the assumption that their asymptotic behaviour at space-like infinity (when considered 
separately) is different. 

As already mentioned in Sec.~\ref{sec:integrability}, the integrability condition 
\eqref{integr_cond} can be used to study the dependence of the ADM mass on the details 
of the gluing construction. After choosing the free parameters entering \eqref{conf_factor} 
and computing the form of $\hat \alpha(x)$ entering the definition of the gluing 
function \eqref{ansatz_beta} in the way described in Sec.~\ref{sec:numer_scheme}, 
the integral \eqref{integr_cond} will be computed numerically using the integrate 
sub-package of the Python SciPy library. The value of the integral computed in this 
way will be denoted by $M_I$ in contrast to the parameter $M$ chosen originally.

Depending on the choice of the free parameters, the right-hand side of \eqref{integr_cond}, 
i.e.~$M_I$, can take on values that do not necessarily agree with $M$. In this case 
the integrability condition would be violated, $\Delta M= M_I - M \neq 0$. Here, we 
will only be interested in the case that $M_I = M$ holds, corresponding to a true 
physical solution. 

\begin{figure}[htb]
 \centering
 \includegraphics[scale = 0.25]{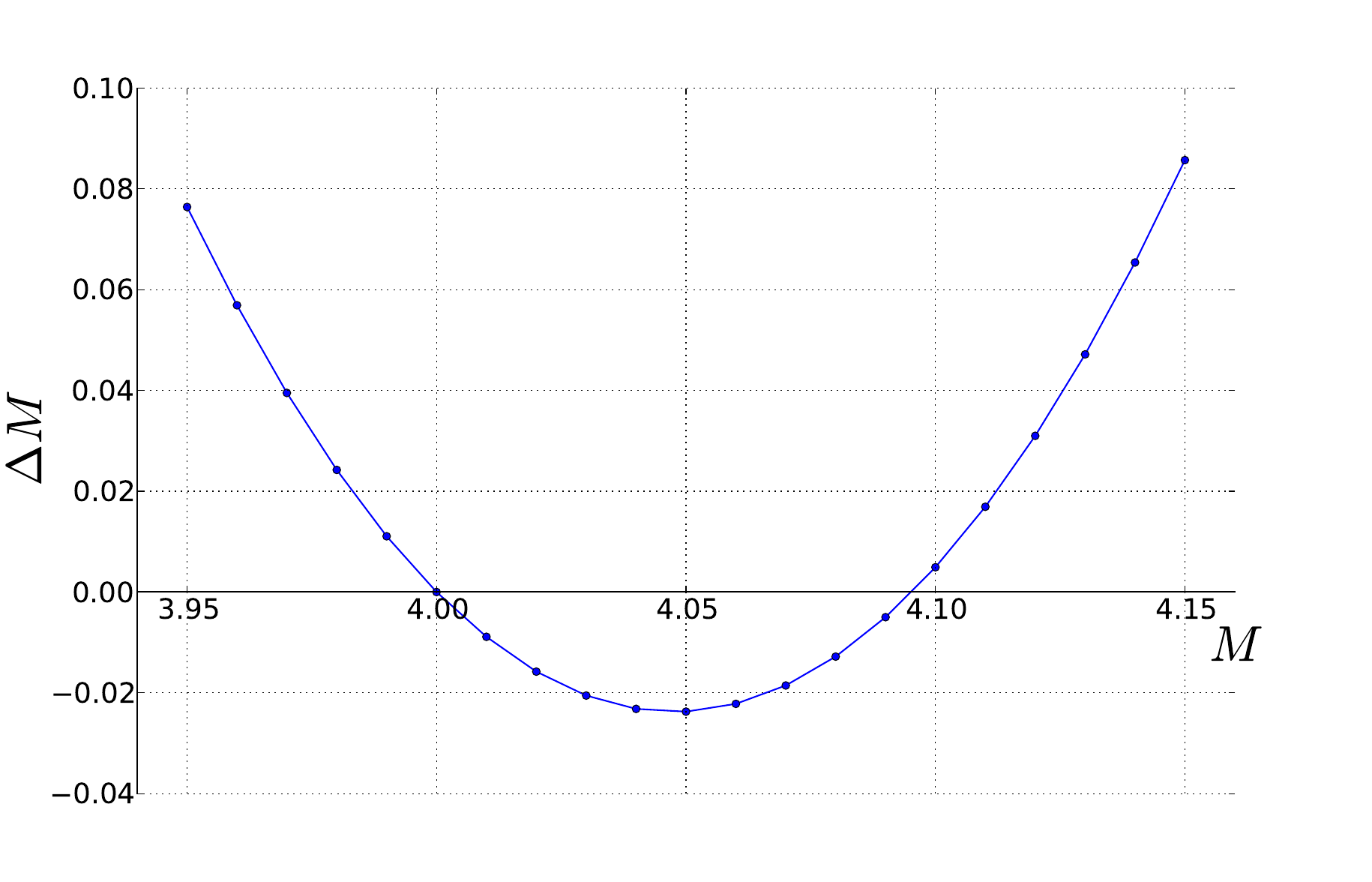}
 \caption{The integrability condition in action. Shown is the difference $\Delta M = M_I - M$ 
 between the integral \eqref{integr_cond} and the originally chosen value $M$ of 
 the ADM mass. As a test case we assume that the data in the interior and exterior 
 of the gluing annulus are Schwarzschildean, i.e. $d = 0$. In this setting, the 
 integrability condition locks the choice of the masses to $M_1 = 4 \pm 3.736 \times 10^{-11}$ 
 or $M_2 \approx 4.095$. The former corresponds to the case of gluing together two 
 identical $M_1 = 2 m$ Schwarzschildean data sets, the latter to the case that the 
 ADM mass of the data in the exterior is larger than the corresponding mass in the 
 interior $M_2 > 2 m$.}
 \label{fig:ADM_mass_Schw}
\end{figure}

To exemplify the use of the condition \eqref{integr_cond}, we will use as a test 
case the scenario that the distance $d$ between the black holes in the interior 
is taken to be zero. In this setting, there is a single black hole of mass $2 m$ 
in the centre to which we attempt to glue a Schwarzschildean end of ADM mass $M$.  
Fig.~\ref{fig:ADM_mass_Schw} depicts how the integrability condition constrains 
the possible choice of the masses $m, M$. Therein, we have plotted the difference 
$\Delta M = M_I - M$ between the integral \eqref{integr_cond} and the originally 
chosen value $M$ of the ADM mass as a function of the ADM mass $M$. For the choice 
$m = 2$, $r_{\mathrm{int}} =100$ and $r_{\mathrm{ext}} = 2\, r_{\mathrm{int}}$ 
the curve crosses the $M$-axis, i.e. the integrability condition $\Delta M = 0$ 
is satisfied, in two distinct points: $M_1 = 4$ and $M_2 \approx 4.095$. The 
first crossing corresponds to the case that the two Schwarzschildean data sets 
are identical $M_1 = 2 m$. Obviously, in this case the Brill wave responsible 
for the gluing must be trivial, i.e. $q = 0$ as it was confirmed at the end of 
Sec.~\ref{sec:results}. The second crossing now corresponds to a setting where 
the two Schwarzschildean data sets we attempt to glue together are different 
$M_2 \neq 2 m$; the Brill wave performing the gluing is now non-trivial, i.e. 
$q \neq 0$. Therefore, the results of Fig.~\ref{fig:ADM_mass_Schw} entail that 
for the class of gluing functions \eqref{ansatz_beta} we consider, the integrability 
condition allows us to glue a Schwarzschildean end of ADM mass $M_1 = 4$ or 
$M_2 \approx 4.095$ to the single black hole of mass $2 m$ residing in the 
centre. Any other combination of the masses would lead to non-physical solutions 
that violate Einstein's equations.

\begin{figure}[htb]
 \centering
  \subfigure[]{
   \includegraphics[scale = 0.24]{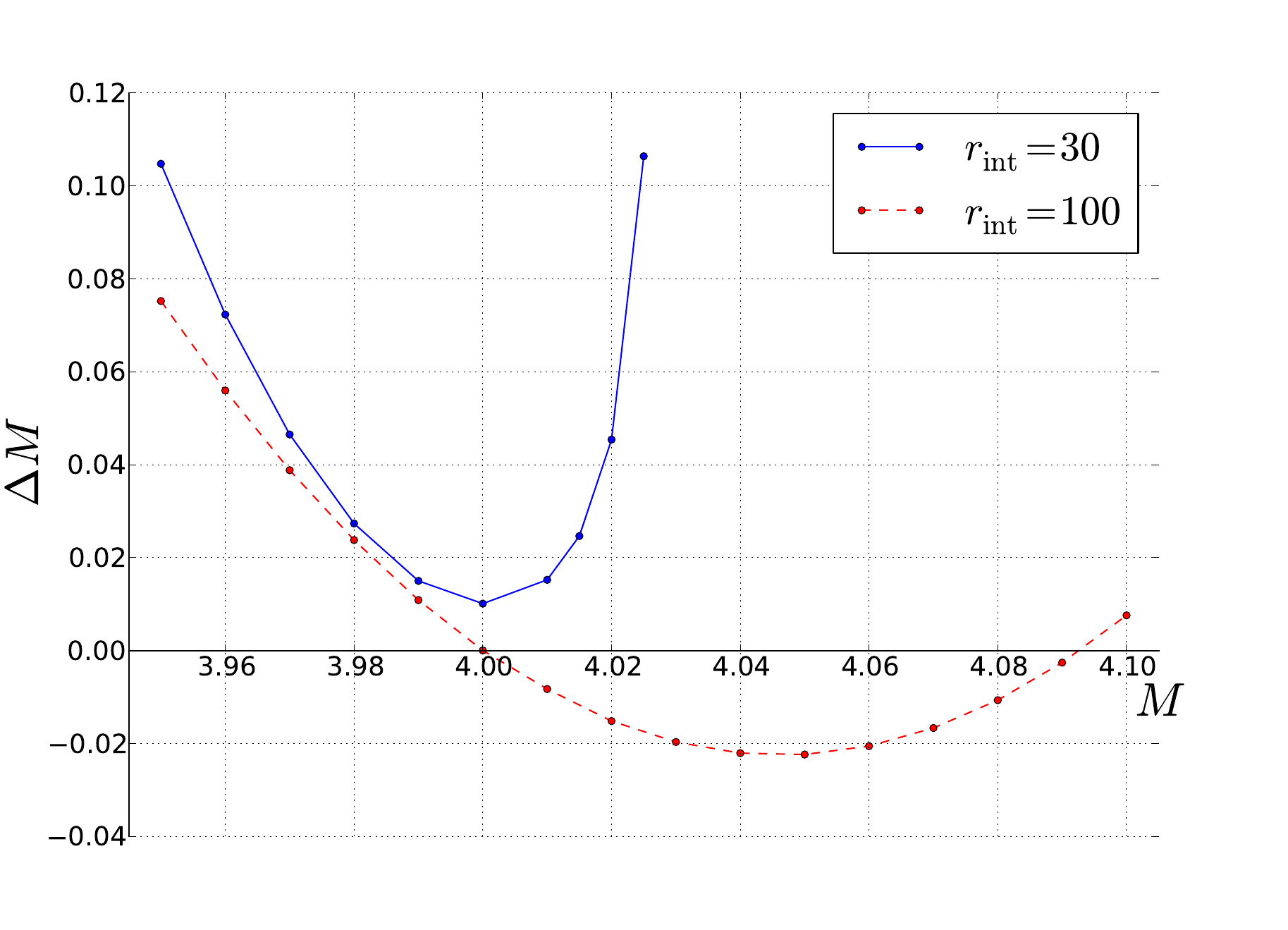}
   \label{fig:ADM_behaviour1}
  }
  \subfigure[]{
   \includegraphics[scale = 0.325]{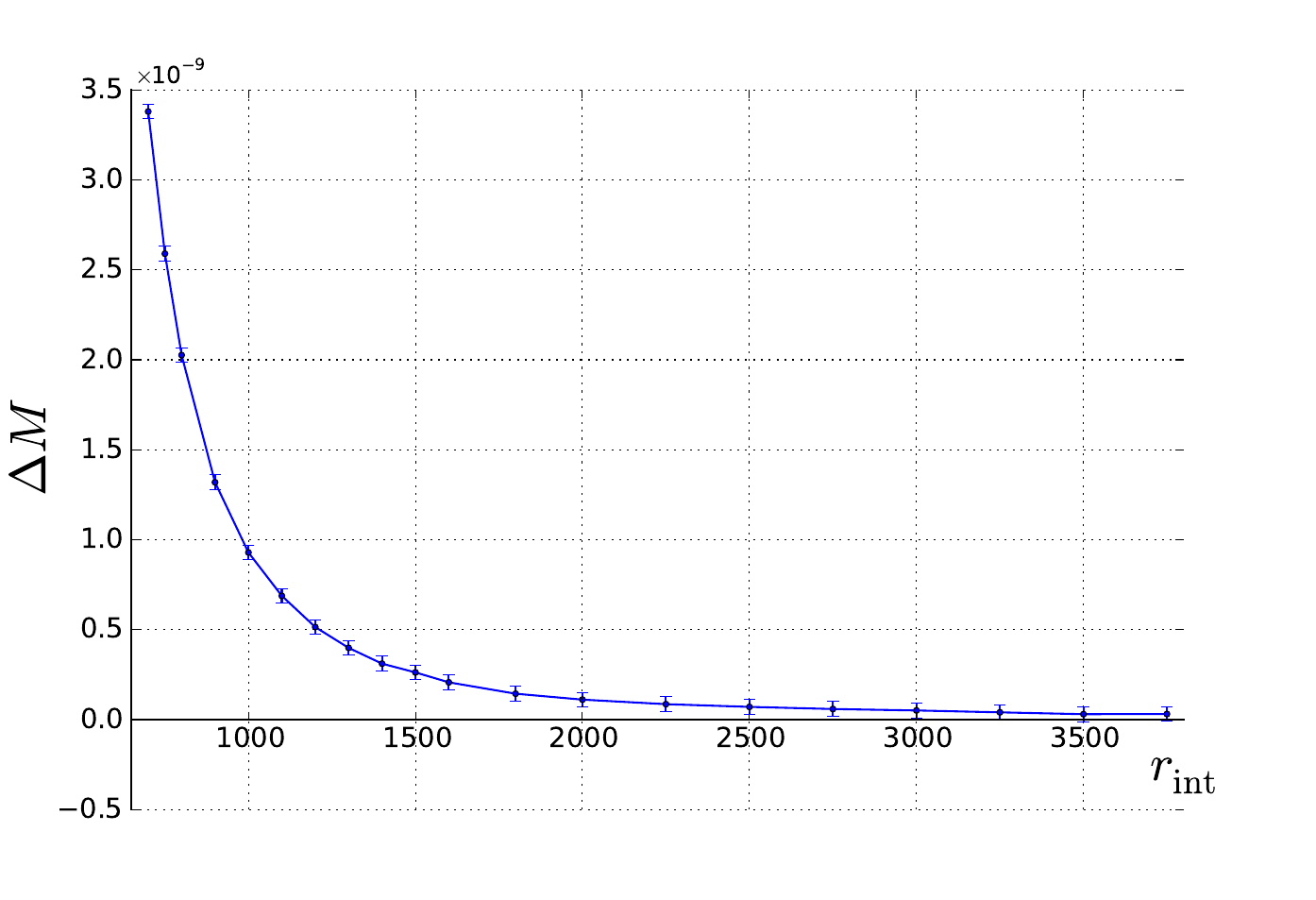}  
   \label{fig:ADM_behaviour2}
  }
 \caption{Behaviour of the ADM mass. (a) The difference  $\Delta M$ between 
 the integral value $M_I$ of the ADM mass and the given parameter $M$ is 
 plotted as a function of the ADM mass $M$ for different values of the gluing 
 radius $r_{\mathrm{int}}$. When placing the gluing annulus at $r_{\mathrm{int}} = 100$ 
 the condition $\Delta M = 0$ is satisfied for $M > 4$; thus only increase 
 of the ADM mass is possible. Notice that in the case $r_{\mathrm{int}} = 30$ 
 the integrability condition cannot be satisfied for any value of $M$; thus 
 \eqref{Poisson_eq} does not have any physically admissible solutions. (b) 
 The difference  $\Delta M$ as a function of the gluing radius $r_{\mathrm{int}}$ 
 for the choice $M = 4$. Whenever $\Delta M > 0$ the ADM mass increases.
 The error bars indicate the numerical error.
 }
 \label{fig:ADM_behaviour}
\end{figure}

Let us return now to the behaviour of the ADM mass for general separations $d$ 
of the two black holes. In order to check if the integrability condition allows 
for a reduction (increase) of the ADM mass, we will fix $m$ and study the dependence 
of the difference $\Delta M = M_I - M$ on the ADM mass $M$ for different locations 
of the gluing annulus. If the violation $\Delta M$ of the integrability condition 
has different signs for two different values of the ADM mass $M$, then according 
to the intermediate value theorem $\Delta M$ must vanish somewhere in between 
these two values of $M$. In Fig.~\ref{fig:ADM_behaviour1}, the free parameters 
were chosen to be $m = 2$, $d = 10$, $r_{\mathrm{ext}} = 2\, r_{\mathrm{int}}$, 
and the gluing annulus has been placed at $r_{\mathrm{int}} = 30$ or $100$. 
The curve for $r_{\mathrm{int}} = 100$ crosses the $M$-axis twice for values 
$M > 2m = 4$---for the first crossing this will be clarified in Fig.~\ref{fig:ADM_behaviour2}---and 
hence the ADM is increased. For $r_{\mathrm{int}} = 30$ the curve does not cross 
the $M$-axis, indicating that, for the choice of the free parameters we are using, 
there are no physically admissible solutions of \eqref{Poisson_eq}. The same 
behaviour is observed for any choice of $r_{\mathrm{int}} \lesssim 40$, implying 
that the gluing is not possible for these positions of the annulus. On the other 
hand, for $r_{\mathrm{int}} \gtrsim 40$ the curve always crosses the $M$-axis 
twice for $M > 4$.

To clarify this point further, we have plotted in Fig.~\ref{fig:ADM_behaviour2} 
for the first crossing the difference $\Delta M$ as a function of the gluing 
radius $r_{\mathrm{int}}$ for \emph{fixed} $M = 4$. (Here, we will concentrate 
on the behaviour of the ADM mass at the first crossing because if the first 
crossing happens for $M_1 > 4$ then certainly the second crossing will happen 
for $M_2 > M_1 > 4$.) Based on the results of Fig.~\ref{fig:ADM_behaviour1}, 
one can safely conclude that close to the first crossing $\Delta M$ decreases 
with $M$; therefore, if $\Delta M$ is positive for $M = 4$ then an appropriate 
increase of $M$ will cause $\Delta M$ to vanish---a setting that leads to 
an increase of the ADM mass of the glued solution. Fig.~\ref{fig:ADM_behaviour2} 
provides strong evidence that the ADM mass is increased for any position of 
the gluing annulus (no matter how far out). For gluing radii larger than 
$r_\mathrm{int} = 3500$ the violation $\Delta M$ becomes of the same order 
of magnitude of the numerical error, i.e. $10^{-11}$,  which indicates that 
it is not possible to draw any decisive conclusions about the behaviour of 
$\Delta M$ there. However, one expects that $\Delta M$ asymptotes to zero 
from positive values as the gluing annulus is progressively placed further 
out: in the limiting case that the gluing is performed at infinity, where 
the two spacetimes become indistinguishable, the Brill wave becomes trivial 
and $\Delta M$ vanishes. 

\begin{figure}[htb]
 \centering
 \includegraphics[scale = 0.36]{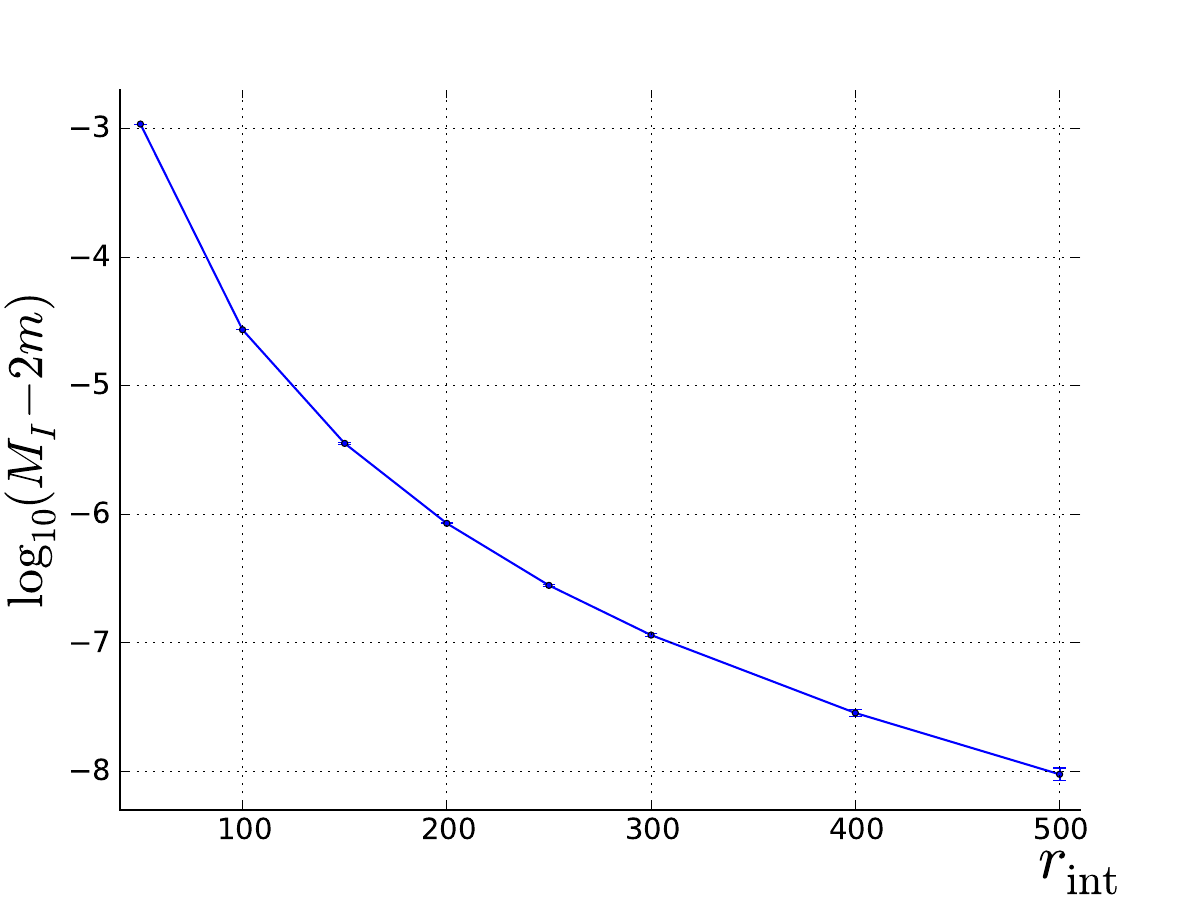}
 \caption{Increase of the ADM mass as a function of the gluing radius. For 
 $r_{\mathrm{int}} = 50$ and $r_{\mathrm{int}} = 100$ the increase amounts 
 to $(1.084 \pm 0.0003)\times 10^{-3}$ and $(2.725 \pm 0.007) \times 10^{-5}$, 
 respectively. By just doubling the gluing radius the increase of the ADM 
 mass decreases by a factor of 40.}
 \label{fig:ADM_increase}
\end{figure}

Let us look a little more closely into the details of the increase of the 
ADM mass and try to determine it quantitatively. As already indicated by 
Fig.~\ref{fig:ADM_behaviour}, the increase is larger for smaller gluing 
radii $r_{\mathrm{int}}$. In Fig.~\ref{fig:ADM_increase} the actual increase 
of the ADM mass, $M_I - 2m$, for different locations of the gluing annulus 
is presented. Notice that the amount of increase, $M_I - 2m$, reduces extremely 
fast to zero with increasing gluing radius: increasing the gluing radius 
from $r_{\mathrm{int}} = 50$ to $100$ results in a decrease of $M_I - 2m$ 
by two orders of magnitude. 

\begin{figure}[htb]
 \centering
 \includegraphics[scale = 0.35]{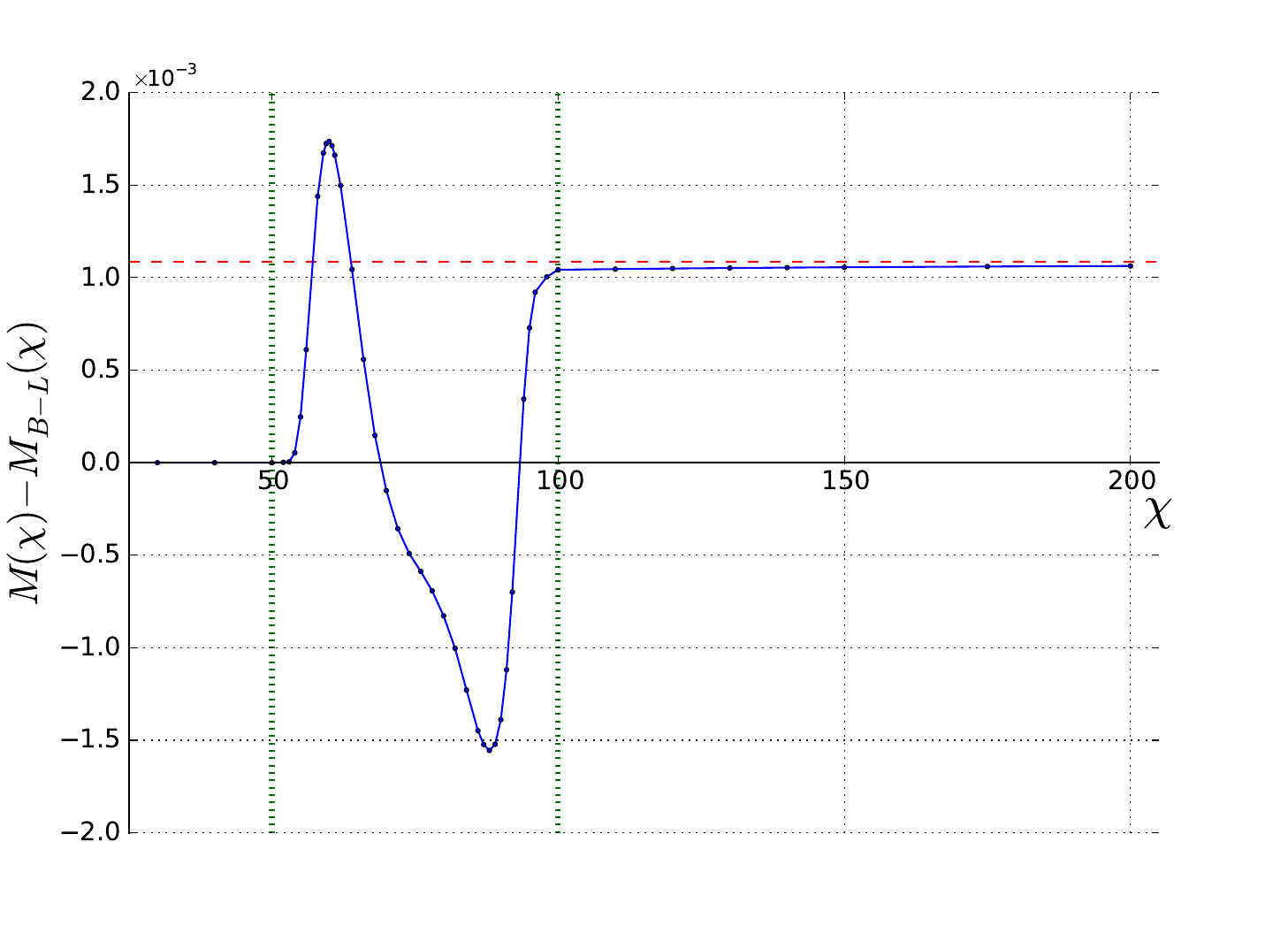}
 \caption{Contribution of the Brill wave to the increase of the ADM mass. 
 The integrability condition \eqref{integr_cond_x} has been used to plot 
 the difference $M(\chi) - M_{B-L}(\chi)$ between glued and pure Brill-Lindquist 
 data as a function of the radial coordinate $\chi$ for the choice $m = 2$, 
 $M = 4.001084$, $d = 10$, $r_{\mathrm{int}} = 50$, $r_{\mathrm{ext}} = 2\, r_{\mathrm{int}}$. 
 The pair of vertical dotted (green) lines denote the span of the gluing 
 annulus. The horizontal dashed (red) line indicates the positive value 
 $M(\infty) - M_{B-L}(\infty) = 0.001084$ given in Fig.~\ref{fig:ADM_increase} 
 corresponding to the amount of increase of the ADM mass. It can be clearly 
 seen that the main contribution to this increase comes from inside the 
 gluing annulus.
}
 \label{fig:ADM_contribution}
\end{figure}

It was mentioned above that the increase of the ADM mass can be attributed 
to the presence of the Brill wave responsible for the gluing. To further 
clarify this point, we will consider the integrability condition \eqref{integr_cond} 
in the form
\begin{equation}
 \label{integr_cond_x}
  M(\chi) = \int^\pi_0 \int^\chi_0 \left[ \left( \frac{1}{\psi}\frac{\partial \psi}{\partial r} \right)^2 + 
  \left( \frac{1}{r\, \psi}\frac{\partial \psi}{\partial \theta} \right)^2 \right] r^2 \sin\theta\, dr\, d\theta,
\end{equation}
where the upper limit of the radial integration takes values in the interval 
$\chi \in [0, \infty)$. Obviously $M(0) = 0$ and in the limit $M(\chi \rightarrow \infty) = M$ 
one obtains the total ADM mass of the gluing construction. In the case we 
have only pure Brill-Lindquist data, i.e.~there is no gluing, we have $M_{B-L}(\chi \rightarrow \infty) = 2\, m$. 
According to Fig.~\ref{fig:ADM_increase}, $M(\infty) - M_{B-L}(\infty)$ is 
always positive. In the interior $\chi \in [0, r_{\mathrm{int}}]$, the difference 
$M(\chi) - M_{B-L}(\chi)$ must be zero as in both cases the data there are 
Brill-Lindquist. Therefore, there must be a point where $M(\chi)$ departs 
from $M_{B-L}(\chi)$ to positive values. This behaviour is studied in Fig.~\ref{fig:ADM_contribution}, 
where the difference $M(\chi) - M_{B-L}(\chi)$ has been plotted as a function 
of $\chi$ for the choice $m = 2$, $M = 4.001084$, $d = 10$, $r_{\mathrm{int}} = 50$, $r_{\mathrm{ext}} = 2\, r_{\mathrm{int}}$ 
corresponding to the numerical solution of Fig.~\ref{fig:num_sol_a}. It is 
apparent that the main contribution to the increase of the ADM mass comes 
from the region where the gluing takes place, i.e. $\chi \in [50, 100]$; 
in the interior $\chi < 50$ the difference $M(\chi) - M_{B-L}(\chi)$ vanishes 
as expected; in the exterior $\chi > 100$ the difference $M(\chi) - M_{B-L}(\chi)$ 
asymptotes to the positive value given in Fig.~\ref{fig:ADM_increase}. Thus, 
it seems that indeed the Brill wave is responsible for the increase of the 
ADM mass. 

\begin{figure}[htb]
 \centering
 \includegraphics[scale = 0.35]{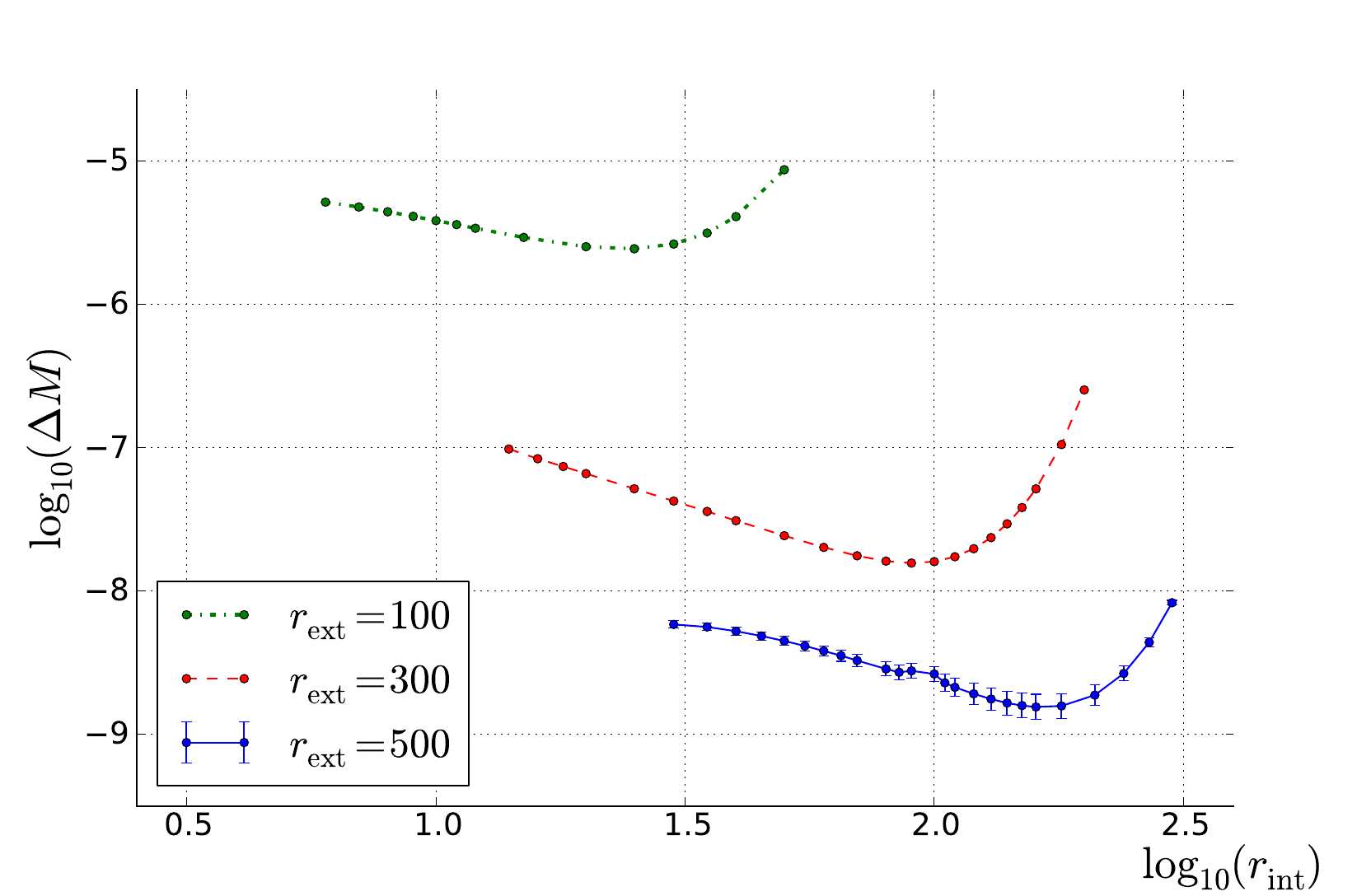}
 \caption{Behaviour of the ADM mass in an extreme scenario. The difference 
 $\Delta M$ between the integral value $M_I$ of the ADM mass and the given 
 parameter $M$ for the configuration $m=2$, $M = 4$, $d=3.2$, and three 
 different locations of the outer boundary $r_\mathrm{ext} = 100, 300, 500$ 
 is presented as a function of the inner gluing radius $r_\mathrm{int}$. 
 (For reasons of presentation we use a logarithmic scale on both axes.) 
 Reduction of the ADM mass is not possible  as $\Delta M$ is always positive.}
 \label{fig:ADM_reduction}
\end{figure}

We conclude with a brief discussion on the possibility of reducing the ADM 
mass. Our extensive numerical study of the solution space of \eqref{Poisson_eq}, 
corresponding to the specific choice \eqref{ansatz_beta} of the gluing function, 
points in the direction that reduction of the ADM mass is not possible. As 
already pointed out in Fig.~\ref{fig:ADM_contribution}, the key point in 
reducing the ADM mass is to find a way to reduce the contribution of the Brill 
wave to it. In Fig.~\ref{fig:ADM_behaviour2} we tried to do so by increasing 
the gluing radius (i.e. placing the gluing annulus further and further out); 
it was shown that reduction of the ADM mass cannot be achieved in this way. 
Other possible ways to ``weaken'' the Brill wave are widening the gluing 
annulus and decreasing the distance between the black holes. In Fig.~\ref{fig:ADM_reduction} 
the behaviour of the ADM mass is studied in a setup where the black holes 
are placed very close to each other and the gluing annulus is extremely wide. 
Specifically, we choose the mass of each one of the black holes to be $m=2$ 
and the distance between them $d=3.2$. For this choice the mass-to-distance 
ratio $m/d = 0.625$ just respects the condition $m/d \lesssim 0.64$, see 
Sec.~\ref{sec:basics}, which prevents the appearance of a third outer horizon 
enclosing both black holes. The horizon of each black hole is $r_\mathrm{hor} = 0.761905$ 
and thus the gluing radius must always satisfy $r_\mathrm{int} \gtrsim 2.4$. 
We fix the mass parameter to be $M=4$. In this setting, we plot in Fig.~\ref{fig:ADM_reduction} 
the difference $\Delta M$ between the integral value $M_I$ of the ADM mass 
and the given parameter $M$ as a function of the position of the inner boundary 
$r_\mathrm{int}$ of the gluing annulus for three different locations of the 
outer boundary: $r_\mathrm{ext} = 100, 300$ and $500$. (Recall that reduction 
or increase of the ADM mass is possible when $\Delta M < 0$ or $\Delta M > 0$, 
respectively.) Our findings indicate that reduction is not possible even in 
this extreme scenario. Although the increase of the ADM mass is smaller the 
further out we place the outer boundary, the behaviour of all curves remains 
qualitatively the same: the difference $\Delta M$ remains always positive 
and an initial decrease of $\Delta M$ is followed by an increase while moving 
the inner boundary towards the outer boundary. The latter behaviour follows 
naturally from the fact that moving the inner boundary towards the outer one 
narrows the gluing annulus, leaving less and less space for the Brill wave to 
perform the gluing and thus increasing its contribution to the ADM mass.

\section{Discussion}
\label{sec:discussion}

The purpose of this paper was to demonstrate for the first time how Corvino's
gluing construction \cite{Corv:2000} can be implemented numerically in order
to compute nontrivial Cauchy data that are Schwarzschild in a neighbourhood 
of space-like infinity.

Our numerical implementation is based on the analytical work by Giulini and 
Holzegel \cite{Giu&Hol:2005}, who applied Corvino's method to axisymmetric
vacuum spacetimes.
In their setting, spacetime is Brill-Lindquist \eqref{B-L_metric} out to some 
radius, is described by a general Brill wave \eqref{Brill_metric} 
along an intermediate gluing region, and is Schwarzschild \eqref{Schw_metric} 
outside this region.
Einstein's equations determine the equation to be solved numerically, 
namely the second-order linear PDE \eqref{Poisson_eq} subject to the boundary 
conditions \eqref{bound_cond_theta} and \eqref{bound_cond_r}. 
In order to obtain physically meaningful solutions, one has to 
constrain the choice of the two mass parameters $m$ and $M$ appearing in the 
definition of the conformal factor. 
It turns out that Einstein's equations imply an integrability 
condition \eqref{integr_cond} that can be used for this purpose. 
In addition, we make sure that the gluing region lies outside of any 
black hole horizons.

To solve numerically the elliptic equation describing the gluing construction, 
we chose to use pseudo-spectral methods.  An extensive convergence analysis, 
both for an artificial exact solution (Sec.~\ref{sec:exact_sol}) and for the 
actual gluing problem (Sec.~\ref{sec:con_analysis}), demonstrates the accuracy 
and convergence of our numerical solutions. Our results confirm the behaviour 
that one would intuitively expect: the numerically computed values of $q$ 
decrease with increasing distance of the gluing annulus from the origin and 
increasing width, see Fig.~\ref{fig:num_solution}. 

Giulini and Holzegel \cite{Giu&Hol:2005} wondered
whether it is possible to choose the gluing parameters in such a way that
the ADM mass $M$ is smaller than $2m$, the sum of the two Brill-Lindquist 
black hole masses. By reducing the ADM mass, one might hope to reduce the 
amount of gravitational radiation that is known to be contained in the 
Brill-Lindquist data \cite{Sperhake2007}. Our findings in Sec.~\ref{sec:reduction} 
suggest that the presence of the Brill wave in the gluing region generically 
tends to increase the ADM mass. We have not been able to reduce the ADM 
mass even in the rather special setup where the black holes are placed 
extremely close to each other and the gluing region extends from close to 
the black hole horizons to a large distance, see Fig.~\ref{fig:ADM_reduction}. 
It should be stressed though that there is a lot of freedom in the choice 
of the gluing function $\beta$. Here we tried only the ansatz \eqref{ansatz_beta}. 
It could be that there exist gluing functions that lead to a reduction of 
the ADM mass, even though we think this is unlikely. So our results do not 
necessarily contradict the asymptotic analysis of \cite{Giu&Hol:2005}. 

We remark that there are other proposals for constructing Cauchy data 
extending to space-like infinity that are not based on Corvino's gluing
method.
For example, Avila \cite{AvilaPhD} considered initial data that are 
only asymptotically static up to a given order at space-like infinity.
It would be interesting to implement this approach numerically as well.
Evolving such data to future null infinity is likely to be more complicated 
than in our approach, where spacetime is known \textit{a priori} in a whole 
neighbourhood of space-like infinity.

Our ultimate goal is to compute an entire spacetime from the Cauchy data 
constructed using the methods described in this paper. As a first step, 
we will evolve our data to a first hyperboloidal surface reaching future 
null infinity; this can then be used as initial data for a hyperboloidal 
evolution code based on either the regular conformal field equations or 
the alternative approaches described in Sec.~\ref{sec:intro}.

\section{Acknowledgments}

We are grateful to Carla Cederbaum, Helmut Friedrich, Domenico Giulini, 
Gustav Holzegel and Mart\'in Reiris for helpful discussions. 
This research is supported by grant RI 2246/2 from the German Research 
Foundation (DFG) and a Heisenberg Fellowship to O.R.


\bibliography{paper}

\end{document}